\documentclass[1p,sort&compress]{elsarticle}
\usepackage{amsmath,amssymb,amscd}
\usepackage{color}
\journal{Nuclear Physics B}
\newcommand{\al}{\alpha}
\newcommand{\be}{\beta}
\newcommand{\de}{\delta}
\newcommand{\ep}{\epsilon}
\newcommand{\vep}{\varepsilon}
\newcommand{\ga}{\gamma}

\newcommand{\la}{\lambda}

\newcommand{\si}{\sigma}
\newcommand{\te}{\theta}
\newcommand{\vt}{\vartheta}
\newcommand{\vp}{\varphi}
\newcommand{\ze}{\zeta}
\newcommand{\La}{\Lambda}
\newcommand{\Si}{\Sigma}
\newcommand{\bphi}{\boldsymbol{\phi}}
\newcommand{\bev}{\mathbf{e}}
\newcommand{\bk}{\mathbf{k}}

\newcommand{\bn}{\mathbf{n}}
\newcommand{\bp}{\mathbf{p}}
\newcommand{\bs}{\mathbf{s}}
\newcommand{\bx}{\mathbf{x}}

\newcommand{\bnu}{{\boldsymbol{\nu}}}

\newcommand{\bt}{{\boldsymbol{\vt}}}
\newcommand{\bte}{{\boldsymbol{\te}}}

\newcommand{\by}{\mathbf{y}}

\newcommand{\tH}{\tilde{H}}
\newcommand{\tK}{\widetilde{K}}

\newcommand{\tS}{\tilde{S}}

\newcommand{\NN}{{\mathbb N}}
\newcommand{\RR}{{\mathbb R}}
\newcommand{\CC}{{\mathbb C}}
\newcommand{\ZZ}{{\mathbb Z}}
\newcommand{\cE}{{\mathcal E}}

\newcommand{\cH}{{\mathcal H}}

\newcommand{\cP}{{\mathcal P}}

\newcommand{\cZ}{{\mathcal Z}}
\newcommand{\fH}{{\mathfrak H}}

\newcommand{\fW}{{\mathfrak W}}
\newcommand\BC{\,\overline{\!C}{}}

\newcommand\Esc{E^{\mathrm{sc}}}

\newcommand\HB{H^{(\mathrm B)}}
\newcommand\HBsc{H^{(\mathrm B)}_{\mathrm{sc}}}
\newcommand\Hsc{H_{\mathrm{sc}}}
\newcommand\tHsc{\tH_{\mathrm{sc}}}

\newcommand\BCB{\overline C^{(\mathrm B)}}
\newcommand\Zsc{Z_{\mathrm{sc}}}
\newcommand\fHsc{\fH_{\mathrm{sc}}}
\newcommand\Lasc{\La_{\mathrm{sc}}}

\newcommand{\pa}{\partial}

\newcommand{\id}{1\hspace{-.25em}{\rm I}}

\newcommand\ket[1]{|#1\rangle}

\newcommand{\ms}{\mspace{1mu}}
\renewcommand{\le}{\leqslant}
\renewcommand{\ge}{\geqslant}

\renewcommand{\geq}{\geqslant}

\newcommand{\card}{\operatorname{card}}

\newcommand{\iu}{\mathrm{i}}
\newcommand{\e}{\mathrm{e}}

\newdefinition{remark}{Remark}
\makeatletter
\def\ps@pprintTitle{%
     \let\@oddhead\@empty
     \let\@evenhead\@empty
     \def\@oddfoot{\footnotesize\itshape\hfill\today}%
     \let\@evenfoot\@oddfoot}

\makeatother
\begin{document}
\begin{frontmatter}
  \title{The exactly solvable spin Sutherland model of $B_N$ type
\\and its related  spin chain}
  \author[BBM]{B. Basu-Mallick}

  \author[UCM]{F. Finkel}

  \author[UCM]{A. Gonz\'alez-L\'opez\corref{cor}}
  \ead{artemio@fis.ucm.es}
  
  \cortext[cor]{Corresponding author}

  \address[BBM]{Theory Group, Saha Institute of Nuclear Physics, 1/AF Bidhan Nagar, Kolkata 700
    064, India}
  \address[UCM]{Departamento de F\'\i sica Te\'orica II, Universidad Complutense, 28040
    Madrid, Spain}
  \begin{abstract}
    We compute the spectrum of the $\mathrm{su}(m)$ spin Sutherland model of $B_N$ type, including
    the exact degeneracy of all energy levels. By studying the large coupling constant limit of
    this model and of its scalar counterpart, we evaluate the partition function of their
    associated spin chain of Haldane--Shastry type in closed form. With the help of the formula
    for the partition function thus obtained we study the chain's spectrum, showing that it cannot
    be obtained as a limiting case of its $BC_N$ counterpart. The structure of the partition
    function also suggests that the spectrum of the Haldane--Shastry spin chain of $B_N$ type is
    equivalent to that of a suitable vertex model, as is the case for its $A_{N-1}$ counterpart,
    and that the density of its eigenvalues is normally distributed when the number of sites $N$
    tends to infinity. We analyze this last conjecture numerically using again the explicit
    formula for the partition function, and check its validity for several values of $N$ and $m$.
  \end{abstract}
\end{frontmatter}

\section{Introduction}\label{sec.intro}

The study of quantum integrable systems with dynamical degrees of freedom exhibiting long-range
interactions had its origin in F. Calogero's celebrated 1971 paper~\cite{Ca71}, where the spectrum
of an $N$-particle system on the line with two-body interactions inversely proportional to the
square of the distance and subject to a confining harmonic potential was exactly computed. An
exactly solvable trigonometric variant of this model was introduced by Sutherland soon
afterwards~\cite{Su71,Su72}. The particles in this model move on a circle,
with two-body interactions proportional to the inverse square of their chord distances. Both of
these integrable models can be substantially generalized by exploiting their
connection with classical root systems, uncovered by Olshanetsky and Perelomov~\cite{OP83}. More
precisely, these authors noted that both the Calogero and Sutherland models are closely
related to the $A_{N-1}$ root system, and constructed generalizations of these models associated
with any (extended) root system.

In a parallel development, Haldane and Shastry independently found an exactly solvable quantum
spin-$\frac{1}{2}$ chain with long-range interactions~\cite{Ha88,Sh88}. The lattice sites of this
su$(2)$ Haldane--Shastry (HS) spin chain are equally spaced on a circle, all spins interacting
with one another through pairwise exchange interactions inversely proportional to the square of
their chord distances. A close relation between the HS chain with su($m$) spin degrees of freedom
and the su($m$) spin version of the Sutherland model~\cite{HH92,HW93,MP93} was subsequently
established using the so-called ``freezing trick''~\cite{Po93,SS93}. More precisely, it was found
that in the strong coupling limit the particles in the spin Sutherland model ``freeze'' at the
coordinates of the equilibrium position of the scalar part of the potential, and the dynamical and
spin degrees of freedom decouple. The equilibrium coordinates coincide with the equally spaced
lattice points of the HS spin chain, so that the decoupled spin degrees of freedom are governed by
the Hamiltonian of the su($m$) HS model. Moreover, in this freezing limit the conserved quantities
of the spin Sutherland model immediately yield those of the HS spin chain, thereby explaining its
complete integrability. The application of the freezing trick to the rational Calogero model with
spin degrees of freedom also led to a new integrable spin chain with long-range
interactions~\cite{Po93}. The sites of this chain ---commonly known in the literature as the
Polychronakos or Polychronakos--Frahm (PF) spin chain--- are unequally spaced on a line, and in
fact coincide with the zeros of the Hermite polynomial of degree $N$~\cite{Fr93}.
The exact partition functions of both the PF and HS spin chains have been exactly
computed by applying the freezing trick~\cite{Po94,FG05}.

Over the years, exactly solvable and integrable one-dimensional quantum many-body systems with
long-range interactions have attracted a great deal of attention in both the physics and the
mathematics literature. In particular, this type of systems have appeared as paradigms of various
condensed matter systems exhibiting generalized exclusion statistics~\cite{Ha96,MS94,Po06}, the
quantum Hall effect~\cite{AI94}, and quantum electric transport phenomena~\cite{BR94,Ca95}. More
recently, quantum integrable spin chains with long-range interactions have played a key role in
calculating higher-loop effects in the spectra of trace operators of planar ${\mathcal N}=4$ super
Yang--Mills theory~\cite{BKS03,Be04,BBL09,Re12}. In the mathematics literature, this type of systems
has been found to be relevant in different fields such as random matrix theory~\cite{TSA95},
multivariate orthogonal polynomials and Dunkl operators~\cite{Fo94,Di97,Du98,FGGRZ01},
and Yangian quantum groups~\cite{BGHP93,Hi95npb,BBHS07,BE08}.

Spin generalizations of the $BC_N$ Calogero--Sutherland model have been extensively studied in the
last few years, and various properties of their related spin chains of HS type have been analyzed
with the help of the freezing trick~\cite{BPS95, Ya95, YT96, CS02, FGGRZ03, EFGR05, BFGR08,
  BFGR09}. Among the other classical root systems, the exceptional ones are comparatively less
interesting in this context, since their associated models consist of at most $8$ particles. On
the other hand, although the $B_N$, $C_N$ and $D_N$ scalar Calogero--Sutherland models have been
studied in the literature~\cite{KPS00,LS04}, their spin versions have been largely ignored.
Recently, however, the $D_N$ spin Calogero~\cite{BFG09} and Sutherland~\cite{BFG11} models, as
well as their associated spin chains, were studied by the present authors and shown to be
nontrivial reductions of their $BC_N$ counterparts.

More precisely, consider the Hamiltonian of the su($m$) spin Sutherland model of $BC_N$
type~\cite{Ya95,EFGR05}
\begin{align}
  H^{(\mathrm{BC})} = -\sum_i \pa_{x_i}^2 &+a\,\sum_{i\neq j}\big[\sin^{-2}
  x_{ij}^-\,(a- \ep\,S_{ij})+\sin^{-2} x_{ij}^+\,(a -\ep\tS_{ij})\big]\notag\\
  &+b\,\sum_i \sin^{-2}\!x_i\,(b-\ep' S_i)+b'\,\sum_i \cos^{-2}\!x_i\, \big(b'-\ep' S_i\big)\,,
  \label{SBC}
\end{align}
where the sums run from $1$ to $N$ (as always hereafter, unless otherwise stated), $a,b,b'>1/2$,
$\ep , \ep' =\pm 1$, and $x_{ij}^{\pm}\equiv x_i\pm x_j$. The operators $S_{ij}$ and $S_i$ in the above
Hamiltonian act on the finite-dimensional Hilbert space
\begin{equation}
  \label{spinbasis}
  \Si=\Big\langle\,|s_1,\dots,s_N\rangle\;\big|\; s_i=-M,-M+1,\dots,M\Big\rangle,
  \qquad M\equiv\frac{m-1}2\in\frac\NN2\,,
\end{equation}
associated with the particles' internal degrees of freedom, as follows:
\begin{equation}
  \begin{aligned}
    & S_{ij}|s_1,\dots,s_i,\dots,s_j,\dots,s_N\rangle=|s_1,\dots,
    s_j,\dots,s_i,\dots,s_N\rangle\,,\\
    & S_i|s_1,\dots,s_i,\dots,s_N\rangle=|s_1,\dots,-s_i,\dots,s_N\rangle\, ,
    \label{SS}
  \end{aligned}
\end{equation}
and we have also used the customary notation $\tS_{ij} =S_iS_jS_{ij}$. If the values of the
coupling constants $b$ and $b'$ in Eq.~(\ref{SBC}) are chosen as indicated in
Table~\ref{tab.Suthmodels}, one obtains su($m$) spin Sutherland models related to the $B_N$, $C_N$
and $D_N$ root systems.

\begin{table}[h]
  \centering
  \begin{tabular}{|l|c|}%
    \hline\vrule height 11pt depth 7pt width0pt
    \hfill Parameters\hfill\hfill& Root system\\\hline
    \vrule height 11pt depth 7pt width0pt $b>\frac{1}{2}$\,, $b'= 0$& $B_N$\\\hline
    \vrule height 11pt depth 7pt width0pt $b=b'>\frac{1}{2}$& $C_N$\\\hline
    \vrule height 11pt depth 7pt width0pt$b=b'=0$& $D_N$\\\hline
  \end{tabular}
  \caption{su$(m)$ Sutherland models of $B_N$, $C_N$, and $D_N$ types}
  \label{tab.Suthmodels}
\end{table}

Since (for instance) the Hamiltonian of the $D_N$-type su($m$) spin Sutherland model is obtained
by formally setting $b=b'=0$ in its $BC_N$ counterpart, one may naively think that all physically
relevant properties of this $D_N$-type model can also be derived from their corresponding $BC_N$
analogs by simply taking the $(b,b')\to 0$ limit. However, the explicit computation of the
spectrum of the model recently performed by the authors shows that this is actually not the
case~\cite{BFG11} (a similar conclusion is reached when comparing the spectra of the $BC_N$ and
$D_N$ Calogero models~\cite{BFG09}.) As a matter of fact, the spectrum of the $D_N$-type spin
Sutherland model cannot be obtained from its $BC_N$ counterpart \cite{EFGR05} through any simple
limiting procedure for the following two reasons. First of all, the Weyl-invariant extended
configuration space of the $D_N$ model ---which turns out to be the $N$-dimensional generalization
of a rhombic dodecahedron--- does not coincide with that of the $BC_N$ model, which is simply a
hypercube. As a consequence, the (scaled) Fourier basis of the Hilbert space of the $BC_N$ model's
auxiliary operator no longer spans a complete set of the Hilbert space of the corresponding
operator for the $D_N$ model. Secondly, while in the $BC_N$ case only one projector of either
positive or negative chirality is needed to construct the Hilbert space of the model from that of
its auxiliary operator, two projectors of $BC_N$ type with opposite chiralities are simultaneously
needed in order to perform a similar construction for the $D_N$ model. Due to these two reasons,
the Hilbert space of the $D_N$ spin model consists of {\em four} ---and not one, as in the case of
the $BC_N$ spin model ---different sectors, characterized by their chirality and parity under
reflections of the particles' coordinates. This fundamental difference explains why the spectrum
of the $D_N$-type spin Sutherland model is essentially different from that of its $BC_N$
counterpart. It also accounts for the greater complexity of the partition function of the
associated chain of $D_N$ type~\cite{BFG11} compared to its $BC_N$ version~\cite{EFGR05}.

Motivated by the nontrivial character of the $D_N$-type models, one can look for other similar
reductions of the $BC_N$-type spin Sutherland model and their related spin chains. From the above
remarks, it is clear that such nontrivial reductions can only be obtained when one or both of the
parameters $b$ and $b'$ are equal to zero, since in that case the singularities of the
Hamiltonian~\eqref{SBC} at $x_i=k\pi$ and/or $x_i=(2k+1)\pi/2$ (with $k\in\ZZ$) disappear, so that
the configuration space of the model differs from that of the general $BC_N$ model. In other
words, the only possible nontrivial reductions of~\eqref{SBC} are the $D_N$ model ($b=b'=0$), the
$B_N$ one ($b>1/2$ and $b'=0$), and the model with $b=0$ and $b'>1/2$. The latter model, which is
not associated with a root system, is nevertheless equivalent to the $B_N$ model under the change
of variables $x_i\mapsto x_i+\frac\pi2$. Thus, apart from the $D_N$ model already studied in
Ref.~\cite{BFG11}, the only new nontrivial reduction of the $BC_N$ Sutherland model is the $B_N$
one. The aim of this paper is precisely to study the $B_N$-type su($m$) spin Sutherland model and
its related spin chain. At the level of the Hamiltonians, the $B_N$-type spin Sutherland model is
also closely related to the $D_N$ one, formally obtained by setting $b=0$ in the $B_N$
Hamiltonian. Nevertheless, our analysis will reveal that the spectrum of the $B_N$-type spin
Sutherland model is essentially different from those of its $BC_N$ and $D_N$ counterparts. It
should also be noted, in this respect, that the Sutherland $C_N$ model is just a trivial reduction
(i.e., a particular instance) of the general $BC_N$ model~\eqref{SBC}, obtained from it by simply
setting $b=b'$.

The organization of this paper is as follows. In Section~\ref{sec.themodels} we define the
Hamiltonians $H^{(\mathrm B)}$ and $\Hsc^{(\mathrm B)}$ of the spin and scalar Sutherland model of
$B_N$ type, respectively. By using the freezing trick, we then construct the Hamiltonian
$\cH^{(\mathrm B)}$ of the associated spin chain of HS type. We show that the sites of this chain,
defined as the coordinates of the (unique) equilibrium point of the scalar part of the spin
Hamiltonian in the principal Weyl alcove of the $B_N$ root system, can be expressed in terms of
the roots of a suitable Jacobi polynomial. Using this characterization, we establish the precise
relations between $\cH^{(\mathrm B)}$ and the Hamiltonians of the HS spin chains associated with
the $BC_N$ and $D_N$ root systems. It turns out that, unlike the case of the corresponding spin
Sutherland models, $\cH^{(\mathrm B)}$ {\em cannot} be obtained from the Hamiltonian of the
$BC_N$-type HS spin chain by taking a suitable limit of its parameters. Section~\ref{sec.spectra}
is devoted to the computation of the spectra of the spin Sutherland model of $B_N$ type and its
scalar version. The main idea behind this computation is relating the Hamiltonians of these models
to an auxiliary scalar operator $H'$, which is a sum of squares of commuting Dunkl operators of
$B_N$ type. Using this property, we explicitly show that $H'$ is upper triangular in the
non-orthonormal basis introduced in Ref.~\cite{BFG11} for the $D_N$ model. In this way one can
compute the spectrum of the operator $H'$, which yields the spectra of both the scalar and spin
Sutherland models of $B_N$ by projecting onto suitable spaces. These results are used in
Section~\ref{sec.partition} to compute the partition function of the HS spin chain of $B_N$ type
as the large coupling limit of the quotient of the partition functions of the spin and scalar
Sutherland models. Using the expression for the partition function of the $B_N$ chain, we compare
its spectrum with those of its $BC_N$ and $D_N$ counterparts for several values of the number of
sites and internal degrees of freedom, verifying in this way that it is not a limiting case of the
latter spectra.

\section{Construction of the $B_N$-type HS spin chain}\label{sec.themodels}

Our main aim in this section is to construct the $B_N$-type HS spin chain from its related spin
Sutherland model by means of the freezing trick. To this end, let us first explicitly write down
the Hamiltonian of the $B_N$-type spin Sutherland model as
\begin{equation}
  H^{(\mathrm B)} = -\sum_i \pa_{x_i}^2 +a\,\sum_{i\neq j}\big[\sin^{-2}
  x_{ij}^-\,(a- \ep\,S_{ij})+\sin^{-2} x_{ij}^+\,(a -\ep\tS_{ij})\big]
  +b\,\sum_i \sin^{-2}\!x_i\,(b-\ep' S_i) \, ,
  \label{SB}
\end{equation}
where $a, b>1/2$ and $\ep , \ep' =\pm 1$. The configuration space $A^{(\mathrm B)}$ of the $B_N$
model~\eqref{SB} is determined by the hard-core singularities of the Hamiltonian on the
hyperplanes $x_i\pm x_j=k\pi$, $x_l=k \pi$ (with $i\ne j$ and $k\in\ZZ$). More precisely, we shall
take $A^{(\mathrm B)}$ as the open subset of $\RR^N$ defined by the inequalities
\begin{equation}\label{defAB}
  0<x_i\pm x_j<\pi\,,~~ 1\le j<i\le N \, ; ~~~~
  0<x_l<\pi\,, ~~ 1\le l\le N \, .
\end{equation}
It is straightforward to check that this set can be equivalently expressed as
\begin{equation}
  \label{defAB1}
  A^{(\mathrm B)}=\{\bx\in\RR^N: 0<x_1<x_2<\cdots <x_N<\pi-x_{N-1}\}\,,
\end{equation}
which is again the principal Weyl alcove of the $B_N$ root system
\begin{equation}\label{BNroots}
  \frac1\pi\,( \pm \bev_i\pm \bev_j)\,, ~~ 1\le i<j\le N\,; ~~~~
  \pm \frac1\pi\, \bev_l\,, ~~ 1\le l\le N\, .
\end{equation}  
Since all wavefunctions and their currents vanish on the boundaries of $A^{(\mathrm B)}$, the
Hamiltonian $H^{(\mathrm B)}$ is naturally defined on some suitable dense subspace of the Hilbert
space $L^2(A^{(\mathrm B)})\otimes\Si$. Let us now put $b=\beta a$ (where $\beta>0$) in
Eq.~\eqref{SB} and decompose $H^{(\mathrm B)}$ into two parts as
\begin{equation}
  \label{dec}
  H^{(\mathrm B)} = \Hsc^{(\mathrm B)}+4a\,h(\bx)\,,
\end{equation}
where
\begin{equation}
  \label{Hsc}
  \Hsc^{(\mathrm B)}=-\sum_i \pa_{x_i}^2 + a(a-1)\,\sum_{i\neq j}\big[\sin^{-2}
  x_{ij}^-\,+\sin^{-2} x_{ij}^+\,\big]
  +\beta a (\beta a -1)\,\sum_i \sin^{-2}\!x_i\,,
\end{equation}
which contains only dynamical degrees of freedom, is the Hamiltonian of the scalar Sutherland
model of $B_N$-type, whereas
\begin{equation}
  h(\bx)=\frac12\sum_{i< j}\big[\sin^{-2} x_{ij}^-\,(1-\,\ep S_{ij})+\sin^{-2}
  x_{ij}^+\,(1-\ep \tS_{ij})\big]\,  
  +\frac{\beta}{4}\,\sum_i \sin^{-2}\!x_i\,(1-\ep' S_i)
  \label{HSB1}
\end{equation}
is a position-dependent multiplication operator featuring the spin degrees of freedom. For the
purposes of applying the freezing trick, we consider the $a\rightarrow \infty$ limit of
$H^{(\mathrm B)}$ (while keeping the value of $\beta$ fixed). The coefficient of the term of order
$a^2$ in the r.h.s. of Eq.~\eqref{dec}, which is given by
\begin{equation}
  \label{UB}
  U^{(\mathrm B)}(\bx)=\sum_{i\neq
    j}\big(\sin^{-2}x_{ij}^-+\sin^{-2}x_{ij}^+\big)+\beta^2
  \,\sum_i \sin^{-2}\!x_i \, , 
\end{equation}
becomes the dominating interaction in this limit. It was shown in Ref.~\cite{CS02} that the scalar
potential $U^{(\mathrm B)}(\bx)$ has a unique minimum $\bt=(\vt_1,\dots,\vt_N)$ in the
configuration space $A^{(\mathrm B)}$. By formally replacing $x_i$ by $\vt_i$ in the r.h.s. of
Eq.\eqref{HSB1}, one obtains the spin chain Hamiltonian
\begin{equation}
  \label{HSB}
  \cH^{(\mathrm B)} = 
  \frac12\sum_{i<j}\Big[\sin^{-2}\vt_{ij}^-\,(1-\, \ep S_{ij})+\sin^{-2}
  \vt_{ij}^+\,(1-\ep\tS_{ij})\Big] 
  +\frac{\beta}{4}\,\sum_i \sin^{-2}\!\vt_i\,(1-\ep' S_i) \, ,
\end{equation}
where $\vt_{ij}^\pm \equiv \vt_i \pm \vt_j$. Now, for sufficiently large $a$ all the eigenfunctions of
$\Hsc^{(\mathrm B)}$ are sharply peaked around the unique minimum $\bt$ of the scalar potential
$U^{(\mathrm B)}$ in the set $A^{(\mathrm B)}$~\cite{Si83}. Hence, if $\vp_i(\bx)$ is an
eigenfunction of $\Hsc^{(\mathrm B)}$ with energy $\Esc_i$ and $\ket{\si_j}$ is an eigenstate of
the chain Hamiltonian $\cH^{(\mathrm B)}$ with eigenvalue $\cE_j$, for $a\gg 1$ we have
\begin{equation}
  h(\bx)\vp_i(\bx)\ket{\si_j}
  \simeq\vp_i(\bx)h(\bt)\ket{\si_j}\equiv\vp_i(\bx)\cH^{(\mathrm B)}\ket{\si_j}
  =\cE_j\vp_i(\bx)\ket{\si_j}\,.
  \label{decp}
\end{equation}
By using Eqs.~\eqref{dec} and \eqref{decp}, it is easy to check that $H^{(\mathrm B)}$ is
approximately diagonal in the basis with elements $\vp_i(\bx)\ket{\si_j}$, and its eigenvalues
$E_{ij}$ satisfy the relation
\begin{equation}
  E_{ij}\simeq \Esc_i+4a\cE_j\,,\qquad a\gg1\,.
  \label{frtr}
\end{equation}
In other words, due to the decoupling of dynamical and spin degrees of freedom in the
$a\rightarrow \infty$ limit, the multiplication operator $h(\bx)$ can be effectively replaced by
$\cH^{(\mathrm B)}$ in Eq.~\eqref{dec}. Consequently, in analogy with the case of other root
systems, it is natural to define the operator $\cH^{(\mathrm B)}$ in Eq.~\eqref{HSB} as the
Hamiltonian of the HS spin chain of $B_N$ type. At first glance, it may seem that one can use
Eq.~\eqref{frtr} to obtain each eigenvalue $\cE_j$ of the chain~\eqref{HSB} in terms of a certain
eigenvalue $E_{ij}$ of the spin Sutherland model of $B_N$ type~\eqref{SB} and a corresponding
eigenvalue $\Esc_i$ of the scalar model~\eqref{Hsc}. In practice, however, the fact that the
eigenvalues $E_{ij}$ and $\Esc_i$ are obviously not independent makes it impossible to use
Eq.~\eqref{frtr} directly to determine the spectrum of the chain~\eqref{HSB} in terms of the
spectra of the Hamiltonians $\HB$ and $\HBsc$. The key idea behind the
freezing trick method pioneered by Polychronakos~\cite{Po94} is to use Eq.~\eqref{frtr} to compute
the chain's partition function. Indeed, the latter equation immediately yields the {\em exact}
identity
\begin{equation}
  \label{ZZZ}
  \cZ(T)=\lim_{a\to\infty}\frac{Z(4aT)}{\Zsc(4aT)}\,,
\end{equation}
expressing the chain's partition function $\cZ$ in terms of the partition functions $Z$ and $\Zsc$
of the Hamiltonians $\HB$ and $\HBsc$, respectively. We shall make use of
this equation in Section~\ref{sec.partition} to explicitly compute
the partition function of the HS spin chain of $B_N$-type.

In the rest of this section we shall discuss the relation of the $B_N$ chain~\eqref{HSB} with
their $BC_N$ and $D_N$ counterparts. To this end, recall~\cite{CS02} that the unique minimum of
the scalar potential $U^{(\mathrm B)}(\bx)$ in the configuration space $A^{(\mathrm B)}$ actually
coincides with the unique maximum in this set of the ground state wave function of the scalar
Hamiltonian~\eqref{Hsc}, given by
\begin{equation}
  \label{varrho}
  \rho(\bx) = \prod_{i<j}\big|\sin x_{ij}^-\,\sin x_{ij}^+\big|^a\,
  \prod_{i}\big|\sin x_i \big|^{\beta a} \, .
\end{equation}
The lattice sites $\vt_i$ of the chain~\eqref{HSB} are thus the unique solution in $A^{(\mathrm
  B)}$ of the nonlinear system of equations:
\begin{equation}
  \label{sitesB}
  \sum_{j;j\ne i}\big(\cot \vt_{ij}^-+\cot \vt_{ij}^+\big)
  +\beta \cot \vt_i \, = \, 0\,,
  \qquad 1\le i\le N\,.
\end{equation}
In order to simplify this system, in analogy with the $BC_N$ and $D_N$ cases
let us define a new set of variables ($\xi_i$) as
\[
\xi_i = \cos(2\vt_i)\,,\qquad 1\le i\le N\,.
\]
Since $\bt\in A^{(\mathrm B)}$, from Eq.~\eqref{defAB1} we obviously have
\[
0<\vt_1<\cdots<\vt_{N-1}<\frac\pi2\,,\qquad 0<\vt_N<\pi\,,
\]
and therefore
\begin{equation}
  \label{zineq}
  1>\xi_1>\xi_2>\cdots >\xi_{N-1}>0\,,\qquad \xi_N<1\,.
\end{equation}
In terms of the variables $(\xi_i)$, the system~\eqref{sitesB} can be rewritten as
\begin{equation}
  \label{sitesz}
  (1-\xi_i^2)\left \{
    \sum_{j;j\ne i}\frac2{\xi_j-\xi_i}+\frac\beta{1-\xi_i}
  \right \}=0\,,\qquad 1\le i\le N\,.
\end{equation}
Let $\xi_{i_0}=\min\{\xi_1,\dots,\xi_N\}=\min\{\xi_{N-1},\xi_N\}$. Since $1-\xi_{i_0}>0$ and
$\xi_j-\xi_{i_0}>0$ for all $j\ne i_0$, the system~\eqref{sitesz} with $i=i_0$ implies that
$\xi_{i_0}=-1$. {}From Eq.~\eqref{zineq} it follows that $i_0=N$. Substituting $\xi_{N}=-1$
into~\eqref{sitesz}, we obtain the following system of equations for the remaining coordinates
$\xi_1,\xi_2,\dots,\xi_{N-1}$:
\begin{equation}
  \label{siteszsimp}
  2(1-\xi_i^2)\sum_{\substack{j=1\\j\ne i}}^{N-1}\frac1{\xi_i-\xi_j}
  =\beta -2 + \xi_i (\beta+2)\,,
  \qquad 1\le i\le N-1\,.
\end{equation}
Comparing~\eqref{siteszsimp} with the system
\begin{equation}
  \label{Jacobi}
  2(1-\ze_i^2)\sum_{\substack{j=1\\j\ne
      i}}^{N}\frac1{\ze_i-\ze_j}=\be-\be'+(\be+\be')\ze_i\,,
  \quad 1\le i\le N\,,
\end{equation}
satisfied by the zeros $\ze_i$ ($i=1,\dots,N$) of the Jacobi polynomial $P^{(\be-1,\be'-1)}_{N}$
(cf.~Ref.~\cite{ABCOP79}), we conclude that the coordinates $\xi_1,\xi_2,\dots, \xi_{N-1}$ are the
zeros of $P_{N-1}^{(\beta - 1,1)}$. In terms of the original site coordinates $\vt_i$ we have
\[
0<\vt_1<\vt_2<\dots<\vt_{N-1}<\vt_N=\frac\pi2\,,
\]
with $P_{N-1}^{(\beta-1,1)}\big(\cos(2\vt_i)\big)=0$ for $i=1,2,\dots,N-1$.

For the purpose of comparing the $B_N$-type HS Hamiltonian \eqref{HSB} with its $BC_N$
counterpart, let us now briefly review the construction of the latter spin chain from the
corresponding spin dynamical model~\cite{EFGR05}. Due to the singularities at the hyperplanes
$x_i\pm x_j=k\pi$, $x_i=k\pi$ and $x_i=\frac\pi2+k\pi$ (with $1\le i<j\le N$ and $k\in\ZZ$), the
configuration space of the spin Sutherland Hamiltonian~\eqref{SBC} can be taken as the principal
Weyl alcove of the $BC_N$ root system
\begin{equation}\label{defABC}
  A^{(\mathrm{BC})} = \Big\{\bx\in\RR^N: 0<x_1<x_2<\cdots<x_N<\frac\pi2\Big\}\,.
\end{equation}
Applying the freezing trick to the Hamiltonian~\eqref{SBC} with $b=\be a$ and $b'=\be' a$ (with
$\be,\be'>0$), one obtains the su($m$) HS spin chain of $BC_N$ type as
\begin{multline}
  \label{HSBC}
  \cH^{(\mathrm{BC})}=\frac12\,\sum_{i<j}\Big[\sin^{-2}\te_{ij}^-\,(1- \ep S_{ij})
  +\sin^{-2}\te_{ij}^+\,(1 - \ep \tS_{ij})\Big]\\
  {}+\frac{\be}{4}\,\sum_i\,\sin^{-2}\te_i (1-\ep' S_i)\, + \frac{\be'}{4}\,\sum_i\,\cos^{-2}\te_i
  (1-\ep' S_i)\, ,
\end{multline}
where $\te_{ij}^\pm \equiv \te_i\pm\te_j$ and $\bte \equiv (\te_1,\ldots,\te_N)$ is the unique
equilibrium in the set $A^{(\mathrm{BC})}$ of the scalar potential
\begin{equation}
  \label{UBC}
  U^{(\mathrm{BC})}(\bx)=\sum_{i\neq
    j}\big(\sin^{-2}x_{ij}^-+\sin^{-2}x_{ij}^+\big)
  +\sum_i\big(\be^2\sin^{-2}x_i+\be'^2\cos^{-2}x_i\big)\,.
\end{equation}
In fact, it is shown in Ref.~\cite{CS02} that the chain sites $\te_i$ can be expressed as
$\ze_i=\cos(2\te_i) $, where $ \ze_i$ are the zeros of the Jacobi polynomial
$P_N^{(\be-1,\be'-1)}$.

Let us now try to find out the precise relation between the $B_N$-type HS spin chain \eqref{HSB}
and the $\be'\to0$ limit of its $BC_N$ counterpart~\eqref{HSBC}. In this context, it should be
noted that the potential $U^{(\mathrm{BC})}$ in Eq.~\eqref{UBC} smoothly reduces to the $B_N$
potential $U^{(\mathrm B)}$ in Eq.~\eqref{UB} in the limit $\be'\rightarrow 0$. Consequently, the
lattice points of the spin chain \eqref{HSBC} should coincide with those of the spin chain
\eqref{HSB} in the $\be'\to0$ limit. In other words, Eq.~\eqref{sitesz} should also yield an
alternative characterization of the coordinates $\xi_i$ as the $N$ roots of the Jacobi polynomial
$P_N^{(\beta-1,-1)}$. This is indeed the case, since Eq.~\eqref{sitesz} obviously reduces
to~\eqref{Jacobi} when $\be'=0$. Alternatively, by using well-known properties of the Jacobi
polynomials we can easily establish the identity
\begin{equation}
  P_N^{(\beta-1,-1)} (z) 
  = \frac{\beta+N-1}{2N} (z+1) P_{N-1}^{(\beta-1,1)} (z)\,,
  \label{ID}
\end{equation}
which confirms the equivalence of both characterizations of the site coordinates of the
$B_N$-type spin chain.

Next, using the identity $\cos^{-2}\te_i=2/(1+\ze_i)$, we can express the last term in the
r.h.s of Eq.~\eqref{HSBC} as
\begin{equation}
  \frac{1}{2}\,\sum_i\frac{\be'}{1+\ze_i} (1-\ep' S_i)\, .
  \label{last}
\end{equation}
Since $\ze_i\to \xi_i$ as $\be'\to0$,
it is clear that the relation $\lim_{\be'\to 0} (1+\ze_i) = 0$ holds only for $i=N$. As a result,
all the terms but the last one in the sum in Eq.~\eqref{last} tend to zero as $\be'\to0$. In order
to evaluate the limit of this last term, we divide~\eqref{Jacobi} by $1+\ze_i$ and sum the
resulting equation over $i$, obtaining
\[
\sum_i\frac{2\be'}{1+\ze_i}=N(\be+\be'+N-1)\,.
\]
Taking the $\be'\to0$ limit of both sides of the above equation, and using the fact that
$\lim_{\be'\to 0} \, \be'/(1+\ze_i) = 0$ for $i \neq N$, we get
\begin{equation}\label{limits}
  \lim_{\be'\to0} \frac{2\be'}{1+\ze_N}=  \, N(N-1 +\beta)\,.
\end{equation}
{}From Eqs.~\eqref{HSBC}, \eqref{last} and~\eqref{limits} it immediately follows that
\begin{equation}\label{impurity}
  \lim_{\be'\to 0}\cH^{(\mathrm{BC})}=\cH^{(\mathrm B)}
  +\frac14\,N(N-1+\beta)\left (1-\ep' S_N \right).
\end{equation}
Thus the $\be'\to0$ of the Hamiltonian of the HS chain of $BC_N$ type yields its $B_N$ analog,
plus an additional term which can be interpreted as an ``impurity'' at the right end of the latter
chain.

Consider now the Hamiltonian of the su($m$) spin Sutherland model of $D_N$ type, which is obtained
by setting $b=b'=0$ in Eq.~\eqref{SBC}:
\begin{equation}
  H^{(\mathrm D)} = -\sum_i \pa_{x_i}^2 + a\,\sum_{i\neq j}\big[\sin^{-2}
  x_{ij}^-\,(a- \ep \,S_{ij})+\sin^{-2} x_{ij}^+\,(a- \ep \tS_{ij})\big]\,.
  \label{SD}
\end{equation}
The configuration space $A^{(\mathrm D)}$ of the $D_N$ model~\eqref{SD} is determined by the
hard-core singularities of the Hamiltonian on the hyperplanes $x_i\pm x_j=k\pi$ (with $i\ne j$ and
$k\in\ZZ$). For $N>2$, it is easy to check that $A^{(\mathrm D)}$ is given by \cite{BFG11}
\begin{equation}
  \label{defAD}
  A^{(\mathrm D)}=\{\bx\in\RR^N:|x_1|<x_2<\cdots<x_N<\pi-x_{N-1}\}\,,
\end{equation}
which is again the principal Weyl alcove of the $D_N$ root system. Application of the freezing
trick to the Hamiltonian \eqref{SD} leads to the Hamiltonian of the su($m$) HS spin chain of $D_N$
type given by
\begin{equation}
  \label{cH}
  \cH^{(\mathrm D)} = \frac12\sum_{i<j}\Big[\sin^{-2}\phi_{ij}^-\,
  (1- \ep \,S_{ij})+\sin^{-2} \phi_{ij}^+\,(1 -\ep \tS_{ij})\Big],
\end{equation}
where the lattice sites $\phi_i$ are the coordinates of the unique minimum $\bphi$ in the set
$A^{(\mathrm D)}$ of the scalar potential
\begin{equation}
  \label{UD}
  U^{(\mathrm D)}(\bx)=\sum_{i\neq
    j}\big(\sin^{-2}x_{ij}^-+\sin^{-2}x_{ij}^+\big).
\end{equation}
Again, defining new variables $\chi_i=\cos(2\phi_i)$, one can show that $\chi_1=-\chi_N=1$ and
that the coordinates $\chi_2,\dots,\chi_{N-1}$ are the zeros of the Jacobi polynomial
$P_{N-2}^{(1,1)}$. Using this characterization, it has been shown that in the $(\be,\be')\to0$
limit the Hamiltonian of the HS chain of $BC_N$ type yields its $D_N$ analog, plus ``impurity''
terms at both ends of the latter chain \cite{BFG11}:
\begin{equation}\label{impurity1}
  \lim_{(\be,\be')\to0}\cH^{(\mathrm{BC})}=\cH^{(\mathrm D)}
  +\frac12\,N(N-1)\Big[1-\frac{\ep'}2(S_1+S_N)\Big].
\end{equation}
Let us now try to establish a relation between the Hamiltonians of the HS spin chains of $B_N$ and
$D_N$ type. To this end, we take the $\be\to0$ limit of both sides of Eq.~\eqref{impurity}, which
yields
\[
\lim_{(\be,\be')\to0} \cH^{(\mathrm{BC})}= \lim_{\be \to0} \cH^{(\mathrm B)} +\frac14\,N(N-1)\left
  (1-\ep' S_N \right).
\]
Comparing the r.h.s. of the above equation with that of Eq.~\eqref{impurity1} we obtain the
relation
\begin{equation}\label{impurity2}
  \lim_{\be \to0} \cH^{(\mathrm B)} = \cH^{(\mathrm D)}
  +\frac14\,N(N-1)\left (1-\ep' S_1 \right),
\end{equation}
which shows that the $\be\to0$ limit of the Hamiltonian of the HS chain of $B_N$ type yields its
$D_N$ analog, plus an ``impurity'' term at the left end of the latter chain.

As mentioned earlier, the Hamiltonians of the $B_N$ and $D_N$ spin Sutherland models can be
obtained from their $BC_N$ counterpart by formally taking some limits of the related coupling
constants. On the other hand, due to the presence of impurity terms in Eqs.~\eqref{impurity},
\eqref{impurity1} and \eqref{impurity2}, it is clear that the Hamiltonians of the HS spin chains
associated with the $B_N$, $D_N$ and $BC_N$ root systems cannot be related to each other by any
simple limiting procedure. Hence, it is natural to expect that the spectrum of the HS spin chain
of $B_N$ type should be qualitatively different from both its $BC_N$ and $D_N$ counterparts. In
this context it should be noted that, in spite of the apparent closeness at the level of their
Hamiltonians, the configuration spaces of the $B_N$, $D_N$ and $BC_N$-type spin Sutherland models
are completely different from each other. Indeed, by comparing Eqs.~\eqref{defAB1}, \eqref{defABC}
and \eqref{defAD} with each other, we find that $A^{(\mathrm D)} \supset A^{(\mathrm B)} \supset
A^{(\mathrm{BC})}$. Since the Hilbert space of a dynamical model is built up from
square-integrable functions defined on the corresponding configuration space, this result clearly
indicates that the spectrum of the spin Sutherland model of $B_N$ type should be qualitatively
different from those of both its $BC_N$ and $D_N$ counterparts.
 
\section{Spectra of the $B_N$-type spin Sutherland model and its scalar
  version}\label{sec.spectra}

In this section we shall compute the spectra of the spin Sutherland model of $B_N$ type~\eqref{SB}
and its associated scalar model~\eqref{Hsc}. We shall employ a well-established
technique~\cite{BGHP93,EFGR05,BFG11}, which is based on relating both of these models to the
auxiliary differential-difference operator
\begin{equation}
  H' = -\sum_i \pa_{x_i}^2 +a\,\sum_{i\neq j}\big[\sin^{-2}
  x_{ij}^-\,(a- K_{ij})+\sin^{-2} x_{ij}^+\,(a -\tK_{ij})\big]
  +b\,\sum_i \sin^{-2}\!x_i\,(b-K_i)\,,
  \label{Hprime}
\end{equation}
where $K_{ij}$ and $K_i$ are coordinate permutation and sign reversing operators, defined by
\begin{align*}
  &(K_{ij}f)(x_1,\dots,x_i,\dots,x_j,\dots,x_N)=f(x_1,\dots,x_j,\dots,x_i,\dots,x_N)\,,\\
  &(K_i f)(x_1,\dots,x_i,\dots,x_N)=f(x_1,\dots,-x_i,\dots,x_N)\,,
\end{align*}
and $\tK_{ij}\equiv K_iK_jK_{ij}$. Let us denote by $\fW$ the group generated by the operators
$K_{ij}$ and $K_i$, i.e., the Weyl group of the $BC_N$ root system, which actually coincides with
  that of the $B_N$ and $C_N$ systems. From Eq.~\eqref{Hprime} it is clear that the operator
  $H'$ is naturally defined on a dense subset of $L^2(C^{(\mathrm B)})$, where $C^{(\mathrm
    B)}\equiv\fW\cdot A^{(\mathrm B)}$. We shall next show that
\begin{equation}\label{CB}
  C^{(\mathrm B)}=\big\{\bx\in\RR^N:0<|x_i\pm x_j|<\pi\,,\enspace x_i\ne0\,;\enspace 1\le i<j\le N\big\}\,.
\end{equation}
Indeed, first of all it is obvious that
\[
\fW\cdot A^{(\mathrm B)}\subset\big\{\bx\in\RR^N:0<|x_i\pm x_j|<\pi\,,\enspace
0<|x_i|<\pi\,;\enspace 1\le i<j\le N\big\}\,.
\]
Adding the two inequalities $-\pi<x_i\pm x_j<\pi$ we immediately obtain $-\pi<x_i<\pi$, so that
\[
\fW\cdot A^{(\mathrm B)}\subset\big\{\bx\in\RR^N:0<|x_i\pm x_j|<\pi\,,\enspace x_i\ne0\,;\enspace
1\le i<j\le N\big\}\equiv C\,.
\]
Hence to prove~\eqref{CB} we need only show that $C\subset\fW\cdot A^{(\mathrm B)}$. To this end,
note that if $\bx\in C$ there is an element $W$ of the Weyl group $\fW$ such that $W\bx=\by$,
where $0<y_1<\cdots<y_N$. Since $C$ is invariant under $\fW$, the vector $\by$ must belong to $C$,
so that $y_{N-1}+y_N<\pi$. Hence $\by\in A^{(\mathrm B)}$, and therefore $\bx=W^{-1}\by\in\fW\cdot
A^{(\mathrm B)}$. This shows that $C\subset \fW\cdot A^{(\mathrm B)}$, thus completing the proof
of Eq.~\eqref{CB}.

As mentioned in the previous section, the operators $\HB$ and $\HBsc$ are naturally defined on
suitable dense subspaces of the Hilbert spaces $L^2(A^{(\mathrm B)})\otimes\Si$ and
$L^2(A^{(\mathrm B)})$, respectively. In order to compute the spectra of these operators, we shall
start by constructing suitable isospectral extensions $\tH$ and $\tHsc$ thereof to appropriate
subspaces of $L^2(C^{(\mathrm B)})\otimes\Si$ and $L^2(C^{(\mathrm B)})$, such that
$\tH=H'\otimes\id$ and $\tHsc=H'$ in the latter subspaces. More precisely, denote by
$\La_{\vep\vep'}$ the projector onto the subspace of $L^2(C^{(\mathrm B)})\otimes\Sigma$
consisting of states with parities $\vep$ and $\vep'$ under particle permutations and simultaneous
reversal of each particle's coordinate and spin, respectively. In other words, the operator
$\La_{\vep\vep'}$ satisfies the relations
\begin{equation}\label{KijSij}
K_{ij}S_{ij}\La_{\vep\vep'}=\vep\, \La_{\vep\vep'}\,,\qquad
K_{i}S_{i}\La_{\vep\vep'}=\vep' \La_{\vep\vep'}\,.
\end{equation}
As shown in Ref.~\cite{BFG11}, there is a natural isomorphism $\enspace\widetilde{}\enspace$
between the spaces $L^2(A^{(\mathrm B)})\otimes\Si$ and $\La_{\vep\vep'}\big(L^2(C^{(\mathrm
  B)})\otimes\Si\big)$, so that $\HB$ is isospectral with the operator $\tilde
H\equiv\widetilde{\phantom a}\circ\HB\circ(\widetilde{\phantom a})^{-1}$ defined on an appropriate
dense subset of the latter space. Similarly, if we denote by $\Lasc$ the projector from
$L^2(C^{(\mathrm B)})$ onto the space of functions symmetric under permutations and even under
sign reversals, defined by the relations
\begin{equation}\label{KiSi}
K_{ij}\Lasc=K_{i}\Lasc=\Lasc\,,
\end{equation}
the spaces $L^2(A^{(\mathrm B)})$ and $\Lasc\big(L^2(C^{(\mathrm B)})\big)$ are again naturally
isomorphic. Hence, denoting (with a slight abuse of notation) this isomorphism by
$\enspace\widetilde{}\enspace$, the operators $\Hsc$ and $\tHsc\equiv\widetilde{\phantom
  a}\circ\Hsc\circ(\widetilde{\phantom a})^{-1}$ are again isospectral. From
Eqs.~\eqref{KijSij}-\eqref{KiSi} and the definition~\eqref{Hprime} of the auxiliary operator $H'$,
it immediately follows that
\begin{equation}\label{tHtHsc}
\tilde H=H'\otimes\id\big|_{\La_{\vep\vep'}(L^2(C^{(\mathrm B)})\otimes\Si)}\,,\qquad
\tHsc=H'\big|_{\Lasc(L^2(C^{(\mathrm B)}))}\,.
\end{equation}

In order to compute the spectra of $\tilde H$ and $\tHsc$, we shall first triangularize the
auxiliary operator $H'$, whose domain is (a dense subset of) the Hilbert space $L^2\big(C^{\mathrm
  (B)}\big)$. In fact, $L^2\big(C^{\mathrm (B)}\big)\equiv L^2\big(\BC^{\mathrm (B)}\big)$, where
$\BC^{\mathrm (B)}$ denotes the closure of the set $C^{\mathrm (B)}$. Using Eq.~\eqref{CB} it is
immediate to show that
\[
\BC^{\mathrm (B)}=\big\{\bx\in\RR^N:|x_i\pm x_j|\le \pi\,,\enspace 1\le i<j\le N\big\}\,,
\]
which coincides with the analogous set for the spin Sutherland model of $D_N$ type studied in
Ref.~\cite{BFG11}. As shown in the latter reference, this set is the $N$-dimensional version of a
rhombic dodecahedron. Furthermore, it was shown in Ref.~\cite{BFG11} that one can construct a
basis of the Hilbert space $L^2\big(\BC^{\mathrm (B)}\big)$ by considering the complex
exponentials $\e^{\iu\sum_j k_j x_j}$ (with $(k_1,\dots,k_N)\in\ZZ^N$) which are periodic on
$\BC^{(\mathrm B)}$, namely the set of functions
\begin{equation}
  \label{fourierB}
  \e^{\iu\ms\sum_j(2n_j+\de) x_j}\,,
  \qquad\bn\equiv(n_1,\dots,n_N)\in\ZZ\,,\quad \de\in\{0,1\}\,.
\end{equation}
By using standard arguments, it can be readily proved that the set of ``gauged'' Fourier functions
\begin{equation}
  \label{Bbasis}
  \vp^{(\de)}_\bn(\bx)\equiv\rho(\bx)\,\e^{\iu\ms\sum_j(2n_j+\de) x_j}\,,
  \qquad\bn\equiv(n_1,\dots,n_N)\in\ZZ\,,\quad \de\in\{0,1\}\,,
\end{equation}
where $\rho$ is defined in Eq.~\eqref{varrho}, is a (non-orthogonal) basis of
$L^2\big(\BC^{\mathrm (B)}\big)$.

\subsection{Triangularization of $H'$}

We shall next define a suitable order in the set~\eqref{Bbasis} so that the action of $H'$ on the
resulting basis is triangular. Note, first of all, that
\begin{equation}\label{V0V1}
  L^2(\BC^{(\mathrm B)})=\fH^{(0)}\oplus\fH^{(1)}\,,
\end{equation}
where $\fH^{(\de)}$ is the closure of the subspace spanned by the basis functions
$\vp^{(\de)}_\bn$ with $\bn\in\ZZ^N$. We will show that $H'$ leaves invariant each of the
subspaces $\fH^{(\de)}$, so that we need only order each subbasis
$\big\{\vp_\bn^{(\de)}\big\}_{\bn\in\ZZ^N}$ in such a way that $H'$ is represented by a triangular
matrix in $\fH^{(\de)}$. To this end, given a multiindex $\bp\equiv(p_1,\dots,p_N)\in\ZZ^N$ we
define
\[ [\bp] = \big(|p_{i_1}|,\dots,|p_{i_N}|\big)\,,\qquad
\text{with}\quad|p_{i_1}|\ge\cdots\ge|p_{i_N}|\,.
\]
If $\bp'\in\ZZ^N$ is another multiindex, we shall write $\bp\prec\bp'$ provided that the first
non-vanishing component of $[\bp']-[\bp]$ is positive. The basis functions
$\big\{\vp_\bn^{(\de)}\big\}_{\bn\in\ZZ^N}$ should then be ordered in any way such that
$\vp_\bn^{(\de)}$ precedes $\vp_{\bn'}^{(\de)}$ whenever $\bnu\prec \bnu'$, where
\begin{equation}\label{bnu}
  \bnu\equiv (2n_1+\de,\dots,2n_N+\de)\,,
\end{equation}
and similarly for $\bnu'$. For instance, $\vp_{(3,1,0)}^{(0)}$ must precede
$\vp_{(2,-3,-1)}^{(0)}$ and $\vp_{(3,1,0)}^{(1)}$ should follow $\vp_{(2,-3,-1)}^{(1)}$, while the
relative precedence of $\vp_{(2,-3,-1)}^{(0)}$ and $\vp_{(1,3,-2)}^{(0)}$ can be arbitrarily
assigned.

In order to compute the action of $H'$ on the basis functions~\eqref{Bbasis}, we shall express the
latter operator in terms of the Dunkl operators of $B_N$ type
\begin{align}
  J_k=\iu\,\pa_{x_k}&+a\sum_{l\neq k}\Big[(1-\iu\cot
  x_{kl}^-)\,K_{kl}+(1-\iu\cot x_{kl}^+)\,\tK_{kl}\Big]\nonumber\\
  &-2a\sum_{l<k}K_{kl}+b(1-\iu\cot x_k)K_k\,,\qquad k=1,\dots,N\,,
  \label{J}
\end{align}
obtained from their $BC_N$ counterparts in Ref.~\cite{EFGR05} by setting $b'=0$. Note that the
natural domain of the operators $J_k$ is the same as that of $H'$, i.e., a suitable dense subspace
of $L^2\big(\BCB\big)$. Setting $b'=0$ in Eq.~(10) of Ref.~\cite{EFGR05} we obtain
\begin{equation}\label{HpJs}
  H'=\sum_k J_k^2\,,
\end{equation}
so that the action of $H'$ on the basis~\eqref{Bbasis} can be deduced from that of the Dunkl
operators~\eqref{J}. In the following discussion, we shall label the basis functions
$\vp_{\bn}^{(\de)}$ simply by $\vp_{\bnu}$, with $\bnu$ defined by~\eqref{bnu}. As in
Ref.~\cite{BFG11}, we shall start by considering the action of $J_k$ on a basis function
$\vp_{\bnu}$ with $\bnu$ nonnegative and nonincreasing. For such a multiindex, we shall use the
notation
\[
\#(s) = \card\{i:\nu_i=s\}\,,\qquad \ell(s) = \min\{i:\nu_i=s\}\,,
\]
with $\ell(s)=+\infty$ if $\nu_i\ne s$ for all $i=1,\dots,N$. For instance, if
$\bnu=(9,7,5,5,3,3)$ then $\#(5)=2$ and $\ell(5)=3$.

Our next step is to prove the key formula
\begin{equation}
  \label{Jifnstruct}
  J_k \vp_{\bnu} = \la_{\bnu,k}\,
  \vp_{\bnu} + \sum_{\substack{\bnu'\in\ZZ^N\\ \bnu'-\bnu\in(2\ZZ)^N,\,\bnu'\!\prec\,\bnu}}
  c_{\bnu,k}^{\bnu'}\,\vp_{\bnu'}\,,
\end{equation}
where $c_{\bnu,k}^{\bnu'}\in\CC$ and
\begin{equation}\label{la}
  \la_{\bnu,k} =
  \begin{cases}
    -\nu_k+2a\big(2\ell(\nu_k)+\#(\nu_k)-k-N-1\big)-b,\quad& \nu_k>0\\[6pt]
    2a(N-k)+b\,,\quad& \nu_k=0\,.
  \end{cases}
\end{equation}
Indeed, using Eq.~\eqref{J}, and performing a lengthy but otherwise straightforward
calculation one finds that
\begin{multline}\label{Jkphinu}
  \frac{J_k\vp_{\bnu}}{\vp_{\bnu}} =-\nu_k-2a(N-1)+2a\sum_{j<k}
  \frac{\al_{jk}^{\nu_j-\nu_k}-1}{\al_{jk}^2-1}
  +2a\sum_{j>k}\frac{\al_{jk}^{\nu_j-\nu_k+2}-1}{\al_{jk}^2-1}\\
  +2a\sum_{j\ne k}
  \frac{\be_{jk}^{2-\nu_j-\nu_k}-1}{\be^2_{jk}-1}
  +2b\,\frac{z_k^{2(1-\nu_k)}-1}{z_k^2-1}-b\,,
\end{multline}
where
\[
\al_{jk}=z_j^{-1}z_k\,,\qquad \be_{jk}=z_jz_k\,,\qquad z_j\equiv\e^{\iu x_j}\,.
\]
Note that all the terms in Eq.~\eqref{Jkphinu} except for the last two also appear in the
corresponding formula for the $D_N$ case, cf.~\cite[Eq.~(51)]{BFG11}. Since
\begin{multline}
\vp_{\bnu}\,\frac{z_k^{2(1-\nu_k)}-1}{z_k^2-1}=-\vp_{\bnu}z_k^{2(1-\nu_k)}\,\frac{1-z_k^{2(\nu_k-1)}}{1-z_k^2}\\
=\begin{cases}
\vp_{\bnu}\,,&\nu_k=0\\
-(z_k^{-2}+\dots+z_k^{-2(\nu_k-1)})\vp_{\bnu}\prec\vp_{\bnu}\,,&\nu_k\ne0\,,
\end{cases}  
\end{multline}
the contribution to $\la_{\bnu,k}$ of the terms proportional to $b$ in Eq.~\eqref{Jkphinu}
is equal to $b(2\de_{\nu_k,0}-1)$. Taking this into account, together with Eqs.~(49)-(50) of
Ref.~\cite{BFG11} for the $D_N$ case, we easily obtain Eqs.~\eqref{Jifnstruct}-\eqref{la}.

Since Eq.~\eqref{Jifnstruct} does not hold in general when $\bnu$ does not belong to
$\big[\ZZ^N\big]$, Eq.~\eqref{la} does not give the complete spectrum of the Dunkl operators
$J_k$. However, in order to compute the spectrum of $H'$ we shall only need the following weaker
result: if $\bnu\in\ZZ^N$ is a multiindex all of whose components have the same parity, then
\begin{equation}
  \label{Jifnarb}
  J_k\vp_{\bnu} =\sum_{\substack{\bnu'\in\ZZ^N\\ \bnu'-\bnu\in(2\ZZ)^N,\,[\bnu']\preceq[\bnu]}}
  \ga_{\bnu,k}^{\bnu'}\,\vp_{\bnu'}
\end{equation}
for some complex constants $\ga_{\bnu,k}^{\bnu'}$. In order to prove this formula, note that if
$\bnu$ is as stated above there is an element $W\in\fW$ such that $\vp_{\bnu}=W\vp_{[\bnu]}$.
Setting $b'=0$ in the commutation relations between the $BC_N$-type Dunkl operators and the
generators of $\fW$ listed in Ref.~\cite{FGGRZ03}, it is straightforward to show that
\[ [J_k,W]=\sum_{j=1}^{2^NN!}c_{jk}W_j\,,\qquad c_{jk}\in\RR\,,
\]
where we have denoted by $W_j$ (with $j=1,\dots,2^NN!$) an arbitrary element of $\fW$. From the
previous equation and the relation $\vp_\bnu=W\vp_{[\bnu]}$ we easily obtain
\[
J_k\vp_{\bnu}=W\big(J_k\vp_{[\bnu]}\big)+\sum_{j=1}^{2^NN!}c_{jk}W_j\vp_{[\bnu]}\,.
\]
Applying Eq.~\eqref{Jifnstruct} to the multiindex $[\bnu]$, and taking into account that the
partial ordering $\prec$ and the parity of the components are invariant under the action of $\fW$,
we easily arrive at Eq.~\eqref{Jifnarb}.\bigskip

We shall next show that the action of $H'$ on each  subbasis
$\big\{\vp_\bn^{(\de)}\big\}_{\bn\in\ZZ^N}$, ordered as explained above, is upper triangular:
\begin{equation}\label{Hpvp}
  H'\vp_\bn^{(\de)}=E_\bn^{(\de)}\vp_\bn^{(\de)}+
  \sum_{\bnu'\prec\bnu}c^{(\de)}_{\bn'\bn}\vp_{\bn'}^{(\de)}\,,\qquad \nu_k\equiv 2n_k+\de\,,\enspace
  \nu'_k\equiv 2n'_k+\de\,,
\end{equation}
where $c^{(\de)}_{\bn'\bn}\in\CC$ and
\begin{equation}\label{Ende}
  E_\bn^{(\de)}=\sum_k\big([\bnu]_k+2a(N-k)+b\big)^2\,.
\end{equation}
Indeed, suppose first that the multiindex $\bnu$ in Eq.~\eqref{Hpvp} is nonnegative and
nonincreasing. Applying $J_k^2$ to both sides of Eq.~\eqref{Jifnstruct} and using Eq.~\eqref{Jifnarb},
it is straightforward to show that
\[
J_k^2\vp_{\bnu}=\la^2_{\bnu,k}\vp_{\bnu}+\sum_{\substack{\bnu'-\bnu\in(2\ZZ)^N\\
    \bnu'\!\prec\,\bnu}} b_{\bnu,k}^{\bnu'}\,\vp_{\bnu'}\,,
\]
with $b_{\bnu,k}^{\bnu'}\in\CC$. From the identity~\eqref{HpJs} we thus obtain
\begin{equation}
  \label{Hpphinu}
  H'\vp_\bnu=\Big(\sum_k\la_{\bnu,k}^2\Big)\vp_{\bnu}
  +\sum_{\substack{\bnu'-\bnu\in(2\ZZ)^N\\ \bnu'\!\prec\,\bnu}}
  \Big(\sum_kb_{\bnu,k}^{\bnu'}\Big)\vp_{\bnu'}\,.
\end{equation}
Suppose, next, that $\bnu\notin[\ZZ]^N$, and let again $W\in\fW$ be such that
$\vp_\bnu=W\vp_{[\bnu]}$. As shown in Ref.~\cite{EFGR05}, the $BC_N$ counterpart of the operator
$H'$ commutes with all the elements of $\fW$. Since $H'$ is obtained from its $BC_N$ analog by
setting $b'=0$, it follows that $[H',W]=0$. Using this fact and applying Eq.~\eqref{Hpphinu} to
$\vp_{[\bnu]}$ we find that
\[
H'\vp_\bnu=W\cdot H'\vp_{[\bnu]}=\Big(\sum_k\la_{[\bnu],k}^2\Big)\vp_{\bnu}
+\sum_{\substack{\bnu'-[\bnu]\in(2\ZZ)^N\\ \bnu'\!\prec\,[\bnu]}}
\Big(\sum_kb_{[\bnu],k}^{\bnu'}\Big)W\vp_{\bnu'}\,,
\]
which establishes~\eqref{Hpvp} with
\begin{equation}\label{Enla}
  E_\bn^{(\de)}=\sum_k\la_{[\bnu],k}^2\,.
\end{equation}
The last step in the proof of Eqs.~\eqref{Hpvp}-\eqref{Ende} is to show that Eq.~\eqref{Enla} can
be simplified to yield Eq.~\eqref{Ende}. For this purpose, let us write $\bp=[\bnu]$ and consider
first the case in which $p_{k-1}>p_k=\cdots=p_{k+r}>p_{k+r+1}\geq 0$. Since $\ell(p_{k+j})=k$ and
$\#(p_{k+j})=r+1$ for $j=0,\ldots,r$, using Eq.~\eqref{la} we obtain
\[
\la_{\bp,k+j}=-p_{k+j}+2a(k+r-j-N)-b= -p_{k+r-j}+2a\big(k+r-j-N)-b,\quad j=0,\ldots,r\,.
\]
Thus in this case we have
\begin{equation}
  \label{partialsum}
  \sum_{l=k}^{k+r}\la_{\bp,l}^2=\sum_{j=k}^{k+r}\big(p_j+2a(N-j)+b\big)^2\,.
\end{equation}
On the other hand, for the case in which $p_{k-1}>p_k=\cdots=p_N=0$ the analog of
Eq.~\eqref{partialsum} follows directly from Eq.~\eqref{la}. Thus Eq.~\eqref{partialsum} is valid
for arbitrary $\bnu\in\ZZ^N$, and Eq.~\eqref{Ende} follows from the latter equation by summing
over $k$.

\subsection{Triangularization of $\HB$ and $\HBsc$}

We shall next make use of the previous results to triangularize $\HB$ and $\HBsc$. As mentioned
above, this problem is equivalent to the triangularization of the extensions $\tH$ and $\tHsc$
acting on their respective Hilbert spaces
$\fH\equiv\La_{\vep\vep'}\big(L^2\big(\BCB\big)\otimes\Si\big)$ and
$\fHsc\equiv\Lasc\big(L^2\big(\BCB\big)\big)$, which can be carried out without difficulty with
the help of Eq.~\eqref{tHtHsc}.

Let us start with the operator $\tH$. By Eq.~\eqref{V0V1}, its Hilbert space can be decomposed as
the direct sum
\begin{equation}\label{fH}
  \fH=\La_{\vep\vep'}\big(\fH^{(0)}\otimes\Si\big)\oplus\La_{\vep\vep'}\big(\fH^{(1)}\otimes\Si\big)\,.
\end{equation}
Let $f(\bx)$ be a function in the domain of $H'$, and let $\ket s\in\Si$ denote an arbitrary spin
state. Since $\tH$ coincides with $H'\otimes\id$ on $\fH$, and the latter operator commutes with
$\La_{\vep\vep'}$ (indeed, it commutes with all the elements of $\fW$), we have
\begin{equation}\label{tHHp}
  \tH\big[\La_{\vep\vep'}\big(f(\bx)\ket s\big)\big]=\La_{\vep\vep'}\big[\big(H'f(\bx)\big)\ket s\big]\,.
\end{equation}
As $H'$ preserves each subspace $\fH^{(\de)}$, the latter equation implies that both subspaces
$\La_{\vep\vep'}\big(\fH^{(\de)}\otimes\Si\big)$ with $\de=0,1$ are invariant under $\tH$. We
shall next verify that $\tH$ acts triangularly on a (non-orthogonal) basis of
$\La_{\vep\vep'}\big(\fH^{(\de)}\otimes\Si\big)$ of the form
\begin{equation}
  \label{psis}
  \psi_{\bn,\bs}^{(\de)}(\bx)=\La_{\vep\vep'}\big(\vp_\bn^{(\de)}(\bx)\ket\bs\big),
\end{equation}
ordered in such a way that $\psi_{\bn,\bs}^{(\de)}$ precedes $\psi_{\bn',\bs'}^{(\de)}$ whenever
$\bnu\prec\bnu'$ (with $\bnu$ defined in~\eqref{bnu}, and similarly $\bnu'$). Since the
functions~$\vp_\bn^{(\de)}$ are a basis of $L^2\big(\BCB\big)$, the spin
wavefunctions~\eqref{psis} are obviously a complete set, but they will not be linearly independent
unless suitable restrictions on the quantum numbers $(\bn,\bs)$ are imposed. More precisely,
the states~\eqref{psis} are a (non-orthogonal) basis of the Hilbert space
$\La_{\vep\vep'}\big(\fH^{(\de)}\otimes\Si\big)$ provided that $\bn\in\ZZ^N$ and
$\bs\in\{-M,-M+1,\dots, M\}^N$ satisfy the following conditions:
\begin{subequations}\label{conds}
  \begin{align}
    &\text{\hphantom{ii}i)}\enspace n_1\geq\cdots\geq n_N\ge0\label{cond1}\\[2mm]
    &\text{\hphantom{i}ii)\enspace If } \de=n_i=0 \text{ then } s_i\geq 0 \text{ for } \vep'=1, \text{ while
    } s_i>0 \text{ for } \vep'=-1.\label{cond2}\\[2mm]
    &\text{iii)\enspace If } n_i=n_j \text{ and } i<j \text{ then }
    \begin{cases}
      s_i\ge s_j\,,& \text{for }\vep=1\\[2mm]
      s_i>s_j\,,& \text{for }\vep=-1      
    \end{cases}
    \label{cond3}
  \end{align}
\end{subequations}
(In condition (ii), it is understood that no additional restriction is imposed on $s_i$ when
either $\de$ or $n_i$ is nonzero).

Indeed, since
\[
\La_{\vep\vep'}(K_{ij}S_{ij})=\vep\La_{\vep\vep'}\,,\qquad
\La_{\vep\vep'}(K_iS_i)=\vep'\La_{\vep\vep'}\,,
\]
acting with suitable operators $K_iS_i$ and $K_{ij}S_{ij}$ on a spin function
$\vp_\bn^{(\de)}(\bx)\ket\bs$ with arbitrary $\bn\in\ZZ^N$ and $\bs$ one can easily show that the
corresponding state $\psi^{(\de)}_{\bn,\bs}$ is either zero or proportional to a
state~\eqref{psis} satisfying the above conditions. (Note, in this respect, that a
state~\eqref{psis} with $\de=n_i=s_i=0$ is symmetric under $(x_i,s_i)\to(-x_i,-s_i)$, and must
therefore vanish identically if $\vep'=-1$.) This shows that the states~\eqref{psis} with
$\bn\in\ZZ^N$ and $\bs$ satisfying the conditions~\eqref{conds} are complete. Their linear
independence is easily checked.

Let us now examine the action of the operator $\tH$ on the basis of
$\La_{\vep\vep'}\big(\fH^{(\de)}\otimes\Si\big)$ given by Eqs.~\eqref{psis}-\eqref{conds}.
From Eqs.~\eqref{Hpvp}-\eqref{Ende} and the identity~\eqref{tHHp} one immediately obtains
\begin{equation}
  \label{tHpsiprev}
  \tH\psi_{\bn,\bs}^{(\de)}=E_{\bn,\bs}^{(\de)}\psi_{\bn,\bs}^{(\de)}+
  \sum_{\bnu'\prec\bnu}c^{(\de)}_{\bn',\bn}\,\psi_{\bn',\bs}^{(\de)}\,,
\end{equation}
where $c^{(\de)}_{\bn',\bn}\in\CC$ and
\begin{equation}\label{Ensde}
  E_{\bn,\bs}^{(\de)}=\sum_k\big(2n_k+\de+2a(N-k)+b\big)^2\,.
\end{equation}
Although the quantum numbers $(\bn',\bs)$ appearing in the r.h.s of Eq.~\eqref{tHpsiprev} need not
satisfy conditions~\eqref{conds}, there is an element $W\in\fW$ such that
$(W\bn',W\bs)\equiv(\bn'',\bs'')$ do satisfy the latter conditions. Since the corresponding state
$\psi_{\bn'',\bs''}^{(\de)}$ differs from $\psi_{\bn',\bs}^{(\de)}$ by at most an overall sign,
and $[\bnu'']=[\bnu']\prec[\bnu]$ implies that $\bnu''\prec\bnu$, it is clear that we can
rewrite~\eqref{tHpsiprev} in the form
\begin{equation}
  \label{tHpsi}
  \tH\psi_{\bn,\bs}^{(\de)}=E_{\bn,\bs}^{(\de)}\psi_{\bn,\bs}^{(\de)}+
  \sum_{\substack{\bn',\bs'\\\bnu'\prec\bnu}}c^{(\de)}_{\bn'\bs',\bn\bs}\,\psi_{\bn',\bs'}^{(\de)}\,,
\end{equation}
for suitable complex constants $c^{(\de)}_{\bn'\bs',\bn\bs}$. Hence the action of $\tH$ on each
subbasis~\eqref{psis}-\eqref{conds} (with fixed $\de\in\{0,1\}$) is indeed triangular, with
eigenvalues given by Eq.~\eqref{Ensde}. The spectrum of $\tH$ is thus obtained from
Eq.~\eqref{Ensde} when $\de=0,1$ and $(\bn,\bs)$ are any quantum numbers satisfying
conditions~\eqref{conds}.

Since, as mentioned at the beginning of this section, the operator $\HB$ is isospectral to $\tH$,
Eq.~\eqref{Ensde} gives the complete spectrum of the spin Sutherland model of $B_N$ type. In
particular, the energies of this model do not depend on the quantum number $\bs$. Therefore, the
degeneracy $d_{\bn}^{(\de)}$ of the eigenvalue~\eqref{Ensde} due to the spin degrees of freedom is
simply the number of basic spin states $\ket\bs$ satisfying conditions~\eqref{conds}. In order to
explicitly compute this degeneracy, let us write the quantum number $\bn$ in the form
\begin{equation}
  \bn=\big(\overbrace{\vphantom{1}p_1,\dots,p_1}^{k_1},\dots,
  \overbrace{\vphantom{1}p_r,\dots,p_r}^{k_r}\big),\qquad p_1>\cdots>p_r\geq0\,.
  \label{neven}
\end{equation}
Using conditions~\eqref{cond2}-\eqref{cond3} we easily find that
\begin{equation}
  \label{dnepsde}
  d_\bn^{(\de)}=
  \begin{cases}\displaystyle
    \binom{m_{\vep\vep'}(k_r)}{k_r}\prod\limits_{i=1}^{r-1}\binom{m_\vep(k_i)}{k_i}\,,\quad& \de=p_r=0\,;\\[5mm]
    \displaystyle \hfill\prod\limits_{i=1}^r\binom{m_\vep(k_i)}{k_i}\,,\hfill& \text{otherwise,}
  \end{cases}
\end{equation}
where
\begin{equation}
  \label{mep}
  m_\vep(k_i)=m+\frac12(1+\vep)(k_i-1)\,,\qquad
  m_{\vep\vep'}(k_r)=\frac12\,\big[m+\vep'\pi(m)+(1+\vep)(k_r-1)\big]
\end{equation}
and $\pi(m)\equiv m \pmod 2$ is the parity of $m$.

Similarly, the spectrum of the scalar Hamiltonian $\tHsc$ can be computed using the fact that it
coincides with $H'$ in the Hilbert space $\fHsc=\Lasc\big(L^2\big(\BCB\big)\big)$, which by
Eq.~\eqref{V0V1} is given by
\begin{equation}
  \label{Hilbertsc}
  \fHsc=\Lasc\big(\fH^{(0)}\big)\oplus\Lasc\big(\fH^{(1)}\big)\,.
\end{equation}
Due to the identity
\[
\tHsc\big(\Lasc f(\bx)\big)=\Lasc\big(H'f(\bx)\big)\,,
\]
it is immediate to show that each of the subspaces $\Lasc\big(\fH^{(\de)}\big)$ is invariant under
$\tHsc$. Just as for the spin model (cf.~Eq.~\eqref{psis}), it can be verified that the functions
\begin{equation}
  \label{psissc}
  \psi_{\bn}^{(\de)}(\bx)=\Lasc\big(\vp_\bn^{(\de)}(\bx)\big),
\end{equation}
where $\bn\in\ZZ^N$ and
\begin{equation}
  \label{nssc}
  n_1\ge\cdots\ge n_N\ge0\,,
\end{equation}
are a (non-orthogonal) basis of $\Lasc\big(\fH^{(\de)}\big)$. Proceeding as above, it is
straightforward to show that the operator $\tHsc$ acts triangularly on the basis obtained by
ordering the set~\eqref{psissc}-\eqref{nssc} in such a way that $\psi_{\bn}^{(\de)}$ precedes
$\psi_{\bn'}^{(\de)}$ whenever $\bnu\prec\bnu'$, and that its eigenvalues $E_{\bn}^{(\de)}$ are
given by the r.h.s of Eq.~\eqref{Ensde}. Since the operators $\HBsc$ and $\tHsc$ are isospectral,
Eq.~\eqref{Ensde} gives the full spectrum of the scalar Sutherland model of $B_N$ type. Of course,
due to the absence of internal degrees of freedom, in this case the degeneracy factors
$d_\bn^{(\de)}$ are equal to one for all quantum numbers $\bn$ and $\de=0,1$.

A remark about the differences between the Hilbert space and spectrum of the spin Sutherland model
of $B_N$ type and those of its $BC_N$ and $D_N$ counterparts is in order at this point. In the
first place, the Hilbert space of the $BC_N$ spin Sutherland model is isomorphic to the subspace
$\La_{\vep\vep'}\big(\fH^{(0)}\otimes\Si\big)$ of $\fH$. On the other hand, the parameter $\vep'$
is not present in the Hamiltonian of the $D_N$ model, but is instead a quantum number which can
take the two values $\pm1$. For this reason, the Hilbert space of the spin Sutherland model of $D_N$
type consists of four sectors, each of which is isomorphic to a space of the form
$\La_{\vep\vep'}\big(\fH^{(\de)}\otimes\Si\big)$ with $\de=0,1$ and $\vep'=\pm1$. Stated
differently, in the $BC_N$ model both parameters $\de$ ($=0$) and $\vep'$ are fixed (and,
therefore, do not appear in the spectrum as quantum numbers), in the $B_N$ model $\de$ is a
quantum number but $\vep'$ is fixed by the Hamiltonian, whereas in the $D_N$ model both $\de$ and
$\vep'$ appear as quantum numbers in the spectrum.

\section{Partition function of the $B_N$-type HS spin chain}\label{sec.partition}

The purpose of this section is to evaluate in closed form the partition function of the
Haldane--Shastry spin chain of $B_N$ type~\eqref{HSB} using the freezing trick. To this end, we
shall make use of the key relation~\eqref{ZZZ} expressing the chain's partition function $\cZ$ in
terms of the partition functions $Z$ and $\Zsc$ of the Hamiltonians $\HB$ and $\HBsc$. In order to
compute the $a\to\infty$ limits of $Z(4aT)$ and $\Zsc(4aT)$, we start by expanding
Eq.~\eqref{Ensde} for the energies of both $\HB$ and $\HBsc$ in powers of $a$, with the result
\begin{equation}
  \label{specHas}
  E_{\bn,\bs}^{(\de)} = a^2E_0+8a\sum_kn_k\Big(N+\frac\be2-k\Big)+2a\,\de N(N+\be-1)+O(1)\,,
\end{equation}
where
\[
E_0=N\be^2+2N(N-1)\be+\frac23\,N(N-1)(2N-1)\,.
\]
Note that, since $a^2E_0$ does not depend on $n_k$, it will clearly not contribute to the quotient
$Z(4aT)/\Zsc(4aT)$. We can therefore subtract this term from the spectra of both $\HB$ and $\HBsc$
for the purposes of computing $\cZ$ through Eq.~\eqref{ZZZ}. With this normalization the
eigenvalues of $\HB$ and $\HBsc$ become $O(a)$ for $a\to\infty$, so that the limits of $Z(4aT)$
and $\Zsc(4aT)$ exist separately. Dropping the term $a^2E_0$ from Eq.~\eqref{specHas} we thus
obtain
\begin{equation}
  \label{ZaT}
  \lim_{a\to\infty}Z(4aT)=\sum_{\de=0,1}\,\sum_{n_1\ge\cdots\ge n_N\ge0}
  d_\bn^{(\de)}\,q^{\frac12\de N(N+\be-1)+2\sum\limits_{i=1}^N n_i(N+\frac\be2-i)}\,,
  \qquad q\equiv\e^{-1/(k_{\mathrm B}T)}\,.
\end{equation}
Using Eq.~\eqref{neven} it can be easily shown that
\begin{equation}\label{psums}
  \sum_{i=1}^N n_i\Big(N+\frac\be2-i\Big)=\sum_{l=1}^rp_l\,\nu_l\,,
\end{equation}
where we have defined
\begin{equation}\label{nulKl}
\nu_l\equiv k_l\Big(N+\frac\be2-K_{l-1}-\frac12(k_l+1)\Big)\,,\qquad K_l\equiv\sum_{i=1}^lk_i\,.
\end{equation}
Substituting Eq.~\eqref{psums} in Eq.~\eqref{ZaT} we have
\begin{align}
  \lim_{a\to\infty}Z(4aT)&=\sum_{\bk\in\cP_N}\sum_{p_1>\cdots>p_r\ge0}
  d_\bn^{(0)}\,q^{2\sum\limits_{l=1}^rp_l\nu_l}+q^{\frac12\,N(N+\be-1)}
  \sum_{\bk\in\cP_N}\sum_{p_1>\cdots>
    p_r\ge0} d_\bn^{(1)}\,q^{2\sum\limits_{l=1}^rp_l\nu_l}\nonumber\\
  \label{Z0Z1}
  &\equiv Z_0(q)+Z_1(q)\,,
\end{align}
where $\cP_N$ is the set of all partitions of the integer $N$ taking order into account,
and we have denoted by $Z_0(q)$ (resp.~$Z_1(q)$) the contribution of the $\de=0$ (resp.~$\de=1$)
sector to $\lim_{a\to\infty}Z(4aT)$.

We shall next proceed to simplify each of the functions $Z_\de(q)$. In the first place, using the
definition of $Z_0(q)$ and the value of the degeneracy factors~$d_\bn^{(\de)}$ in
Eq.~\eqref{dnepsde} we easily arrive at the formula
\begin{equation}\label{Z0q}
Z_0(q)=\sum_{\bk\in\cP_N}d_0d_1\sum_{p_1>\cdots>
  p_r>0}q^{2\sum\limits_{l=1}^rp_l\nu_l}
+\sum_{\bk\in\cP_N}d_0d_2\sum_{p_1>\cdots> p_{r-1}>0}
q^{2\sum\limits_{l=1}^{r-1}p_l\nu_l}\,,
\end{equation}
where we have set
\[
d_0=\prod_{i=1}^{r-1}\binom{m_\vep(k_i)}{k_i}\,,\qquad
d_1=\binom{m_\vep(k_r)}{k_r}\,,\qquad
d_2=\binom{m_{\vep\vep'}(k_r)}{k_r}\,.
\]
Proceeding as in Ref.~\cite{EFGR05} it is straightforward to obtain the key identity
\begin{equation}\label{keyid}
\sum_{p_1>\cdots>
  p_s>0}q^{2\sum\limits_{l=1}^rp_l\nu_l}
=\prod_{i=1}^s\frac{q^{\cE(K_i)}}{1-q^{\cE(K_i)}}\,,
\end{equation}
where the dispersion relation $\cE(t)$ is defined by
\begin{equation}
  \cE(t)=t(2N+\be-1-t)\,.
\end{equation}
Substituting this identity with $s=r$ and $s=r-1$ in Eq.~\eqref{Z0q} we find that
\begin{equation}\label{Z0final}
Z_0(q)=\sum_{\bk\in\cP_N}d_0d_1\prod_{i=1}^r\frac{q^{\cE(K_i)}}{1-q^{\cE(K_i)}}
+\sum_{\bk\in\cP_N}d_0d_2\prod_{i=1}^{r-1}\frac{q^{\cE(K_i)}}{1-q^{\cE(K_i)}}\,.
\end{equation}
Consider now the function $Z_1(q)$, explicitly given by
\[
Z_1(q)=q^{\frac12N(N+\be-1)}\sum_{\bk\in\cP_N}d_0d_1\sum_{p_1>\cdots>
  p_r\ge 0}q^{2\sum\limits_{l=1}^rp_l\nu_l}
\]
This formula can be simplified by using the identity
\[
\sum_{p_1>\cdots>
  p_r\ge0}q^{2\sum\limits_{l=1}^rp_l\nu_l}
=\prod_{i=1}^{r-1}\frac{q^{\cE(K_i)}}{1-q^{\cE(K_i)}}\cdot \frac{1}{1-q^{\cE(K_r)}}\,,
\]
which easily follows from Eq.~\eqref{keyid}, with the result
\begin{equation}\label{Z1q}
  Z_1(q)=q^{\frac12N(N+\be-1)}\sum_{\bk\in\cP_N}d_0d_1
  \prod_{i=1}^{r-1}\frac{q^{\cE(K_i)}}{1-q^{\cE(K_i)}}\cdot \frac{1}{1-q^{\cE(K_r)}}\,.
\end{equation}
Note that $K_r=N$, so that
\[
\cE(K_r)=N(N+\be-1)\,.
\]
Substituting Eqs.~\eqref{Z0final} and \eqref{Z1q} in Eq.~\eqref{Z0Z1} we obtain
\begin{align*}
\lim_{a\to\infty}Z(4aT)&=\sum_{\bk\in\cP_N}d_0d_1\prod_{i=1}^r\frac{q^{\cE(K_i)}}{1-q^{\cE(K_i)}}
+\sum_{\bk\in\cP_N}d_0d_2\prod_{i=1}^{r-1}\frac{q^{\cE(K_i)}}{1-q^{\cE(K_i)}}\\
&\quad+q^{\frac12N(N+\be-1)}\sum_{\bk\in\cP_N}d_0d_1
\prod_{i=1}^{r-1}\frac{q^{\cE(K_i)}}{1-q^{\cE(K_i)}}\cdot \frac{1}{1-q^{\cE(K_r)}}\\
&=\sum_{\bk\in\cP_N}\prod_{i=1}^{r-1}q^{\cE(K_i)}\cdot\prod_{i=1}^r\frac{1}{1-q^{\cE(K_i)}}
\cdot d_0\bigg[
d_1\big(q^{\frac12\cE(K_r)}+q^{\cE(K_r)}\big)+d_2(1-q^{\cE(K_r)})
\bigg]\,.
\end{align*}
After a straightforward simplification, this equation yields the following explicit formula for
the $a\to\infty$ limit of the partition function of the spin Sutherland model of $B_N$ type:
\begin{multline}\label{Z4aTfinal}
  \lim_{a\to\infty}Z(4aT)\\
  =\big(1+q^{\frac{N}2(N+\be-1)}\big)\sum_{\bk\in\cP_N}
  \prod_{i=1}^{r-1}q^{\cE(K_i)}\cdot\prod_{i=1}^r\frac{1}{1-q^{\cE(K_i)}}\,
  \cdot\, d_0
\bigg[d_1q^{\frac{N}2(N+\be-1)}+d_2\big(1-q^{\frac{N}2(N+\be-1)}\big)\bigg].
\end{multline}

We shall next evaluate the partition function of the scalar Sutherland model of $B_N$ type in the
limit $a\to\infty$. As mentioned above, the energies of this model are still given by the r.h.s.
of Eq.~\eqref{specHas}, although in this case there is no degeneracy due to the spin degrees of
freedom. Thus the large $a$ limit of the partition function $\Zsc(4aT)$ is given by
Eq.~\eqref{ZaT} with $d_\bn^{(\de)}=1$:
\begin{align}
  \label{ZscaT}
  \lim_{a\to\infty}\Zsc(4aT)&=\sum_{\de=0,1}\,\sum_{n_1\ge\cdots\ge n_N\ge0}
  q^{\frac12\de N(N+\be-1)+2\sum\limits_{i=1}^N n_i(N+\frac\be2-i)}\nonumber\\
  &= \big(1+q^{\frac12 N(N+\be-1)}\big)\sum_{n_1\ge\cdots\ge n_N\ge0}
  q^{2\sum\limits_{i=1}^N n_i(N+\frac\be2-i)}
  \,.
\end{align}
Evaluating the last sum as in~Ref.~\cite{EFGR05} we readily obtain
\begin{equation}\label{Zsc4aTfinal}
 \lim_{a\to\infty}\Zsc(4aT)=\big(1+q^{\frac12 N(N+\be-1)}\big)\prod_{i=1}^N\big(1-q^{\cE(i)}\big)^{-1}\,.
\end{equation}

The partition function of the Haldane--Shastry spin chain of $B_N$ type~\eqref{HSB} is easily
computed by inserting Eqs.~\eqref{Z4aTfinal} and~\eqref{Zsc4aTfinal} into the key
relation~\eqref{ZZZ}. In order to simplify the resulting expression, we define $N-r$
integers $K_1'<\cdots<K_{N-r}'$ in the range $1,\dots,N-1$ by
\[
\big\{K_1',\dots,K_{N-r}'\big\}=\big\{1,\dots,N-1\big\}-\{K_1,\dots,
K_{r-1}\big\}\,.
\]
Using this notation, we finally arrive at the following closed-form expression for the partition
function of the spin chain~\eqref{HSB}:
\begin{equation}
  \label{ZBN}
  \cZ=\sum_{\bk\in\cP_N}
  \prod_{i=1}^{r-1}q^{\cE(K_i)}\cdot\prod_{j=1}^{N-r}\big(1-q^{\cE(K'_j)}\big)\,
  \cdot\, d_0
\bigg[d_1q^{\frac{N}2(N+\be-1)}+d_2\big(1-q^{\frac{N}2(N+\be-1)}\big)\bigg].
\end{equation}
In particular, from the latter equation it is clear that $\cZ$ is a finite sum of powers
of $q$, as should be the case for a finite system.

\begin{figure}[h]
  \centering
  \includegraphics[width=.48\textwidth]{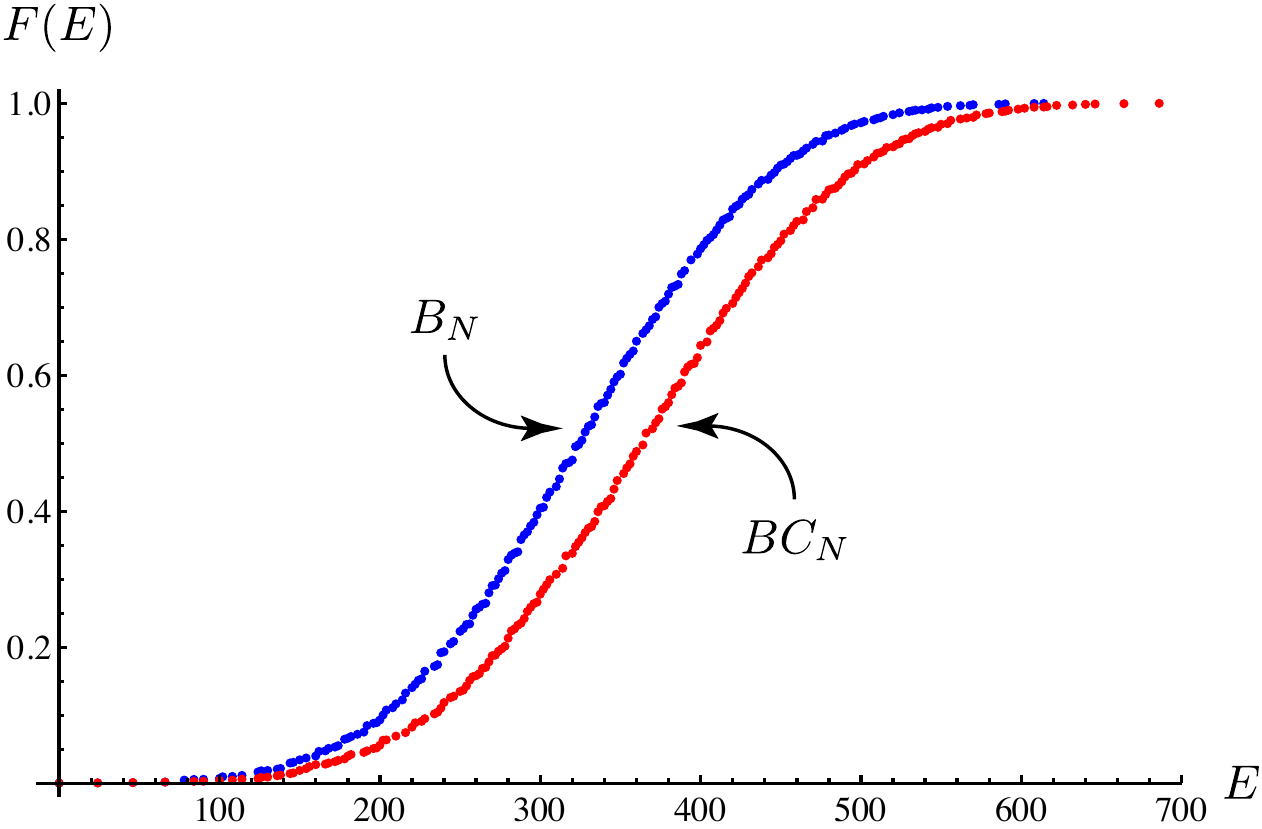}\hfill
  \includegraphics[width=.48\textwidth]{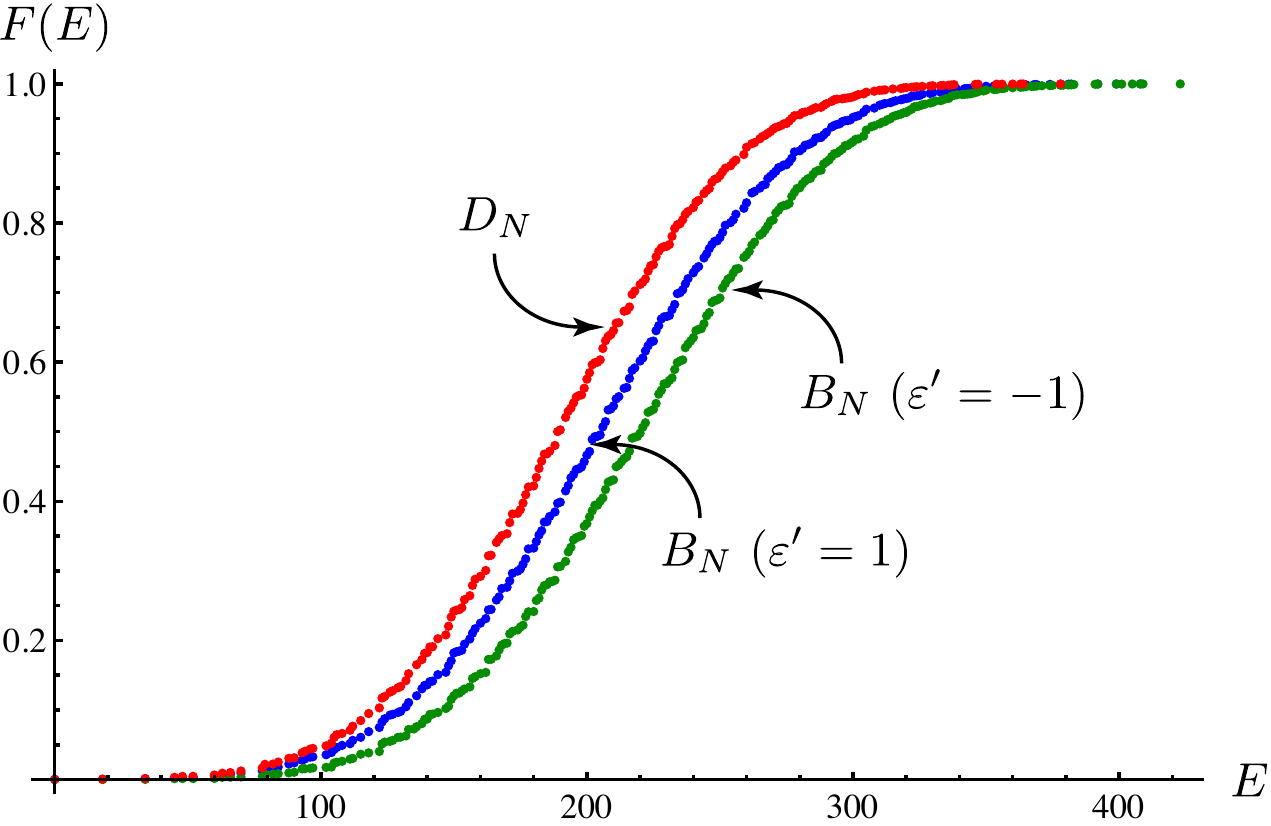}
  
  \caption{Left: cumulative level density of the ferromagnetic HS chain of $B_N$ type with $N=12$,
    $m=2$, and $\be=2$ (blue) vs. its $BC_N$ counterpart with $\be+\be'=2$ (red). Right:
    cumulative level density of the ferromagnetic HS chain of $D_N$ type with $N=10$ and $m=3$
    (red) compared to its $B_N$ analogs with $\be\to0$ and $\vep'=1$ (blue), $\vep'=-1$ (green).
    Note that, by Eq.~\eqref{mep}, the energies of both the $B_N$ and the $BC_N$ chains are
    independent of $\vep'$ when $m$ is even, while the spectrum of the $BC_N$ chain depends on
    $\be$ and $\be'$ through the combination $\be+\be'$ on account of Eq.~\eqref{ZBCN}.}
\label{fig.chains}
\end{figure}

For comparison purposes, we note that the partition function $\cZ^{(\mathrm{BC})}$ of the HS spin
chain of $BC_N$ type~\eqref{HSBC}, which can be inferred from Eqs.~(52)-(54) in
Ref.~\cite{EFGR05}, may be written in the notation of the present paper as
\begin{equation}\label{ZBCN}
\cZ^{(\mathrm{BC})}=\sum_{\bk\in\cP_N}
\prod_{i=1}^{r-1}q^{\tilde\cE(K_i)}\cdot\prod_{j=1}^{N-r}\big(1-q^{\tilde\cE(K'_j)}\big)\,
\cdot\, d_0 \bigg[d_1q^{N(N+\be+\be'-1)}+d_2\big(1-q^{N(N+\be+\be'-1)}\big)\bigg],
\end{equation}
with $\tilde\cE(t)\equiv t(2N+\be+\be'-1-t)$. Comparing Eqs.~\eqref{ZBN} and~\eqref{ZBCN}, it is
apparent that the partition function $\cZ^{(\mathrm{BC})}$ does \emph{not} tend to its $B_N$
analog $\cZ$ in the limit $\be'\to0$. Likewise, it is clear that the expression of the partition
function of the spin Sutherland model of $D_N$ type given by Eqs.~(92) and (95) in
Ref.~\cite{BFG11} is much more complex in nature than its $B_N$ counterpart~\eqref{ZBN} with
$\be=0$. Indeed, the fact that the partition functions of the $BC_N$, $B_N$ and $D_N$ models
cannot be obtained from one another by taking appropriate limits of the parameters $\be$ and
$\be'$ is in agreement with the presence of boundary terms in
Eqs.~\eqref{impurity1}-\eqref{impurity2}. In order to illustrate this remark, in
Fig.~\ref{fig.chains} we have compared the spectra of the $B_N$ chain with its $BC_N$ and $D_N$
counterparts for different choices of $N$, $m$, and $\be$. More precisely, in the latter figure we
have plotted the (normalized) cumulative level density of these chains, defined by
\[
F(E)=\frac1{m^N}\sum_{E_i\le E}\de_i\,,
\]
where $E_1<\cdots<E_n$ are the distinct energies and $\de_i$ denotes the degeneracy of the energy
$E_i$. It is apparent from these and similar plots that the spectra of the $B_N$, $BC_N$ and $D_N$
chains cannot be obtained from one another by taking appropriate limits of the parameters $\be$
and $\be'$.

\begin{figure}[h]
  \centering
  \includegraphics[width=.48\textwidth]{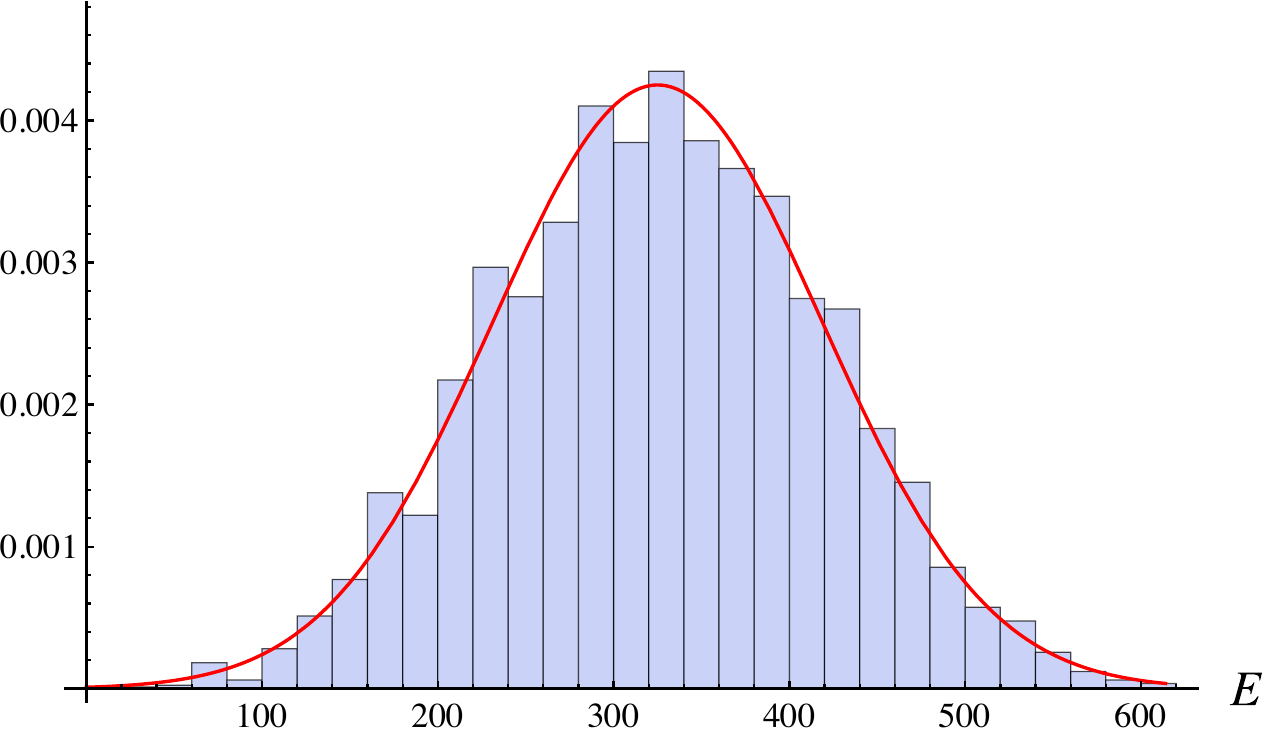}\hfill
  \includegraphics[width=.48\textwidth]{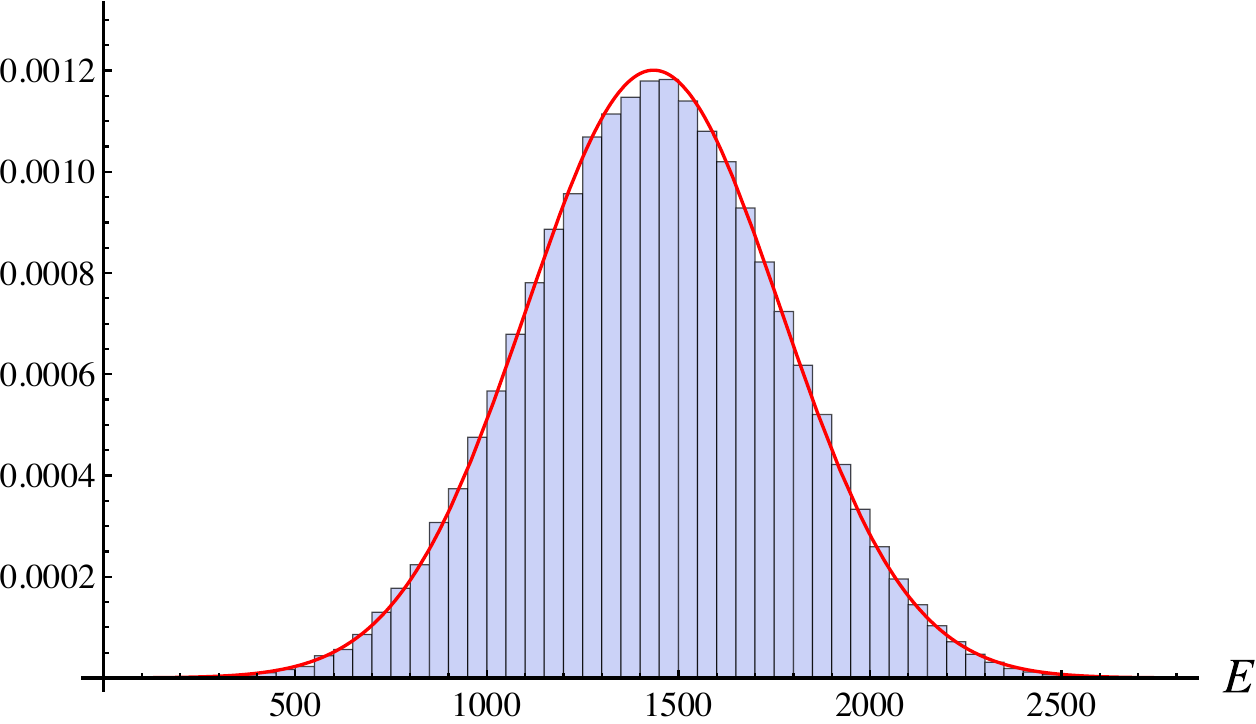}
  
  \caption{Probability density function histogram of the level density of an su(2) ferromagnetic
    HS chain of $BC_N$ type with parameter $\be=2$ for $N=12$ spins (left) and $N=20$ spins
    (right), compared to a normal distribution with the same mean and standard deviation as the
    spectrum (continuous red line).}
  \label{fig.gauss}
\end{figure}

On the other hand, the obvious structural similarity between Eqs.~\eqref{ZBN} and~\eqref{ZBCN} and
the fact\footnote{J.C. Barba, F. Finkel, A. Gonz\'alez-L\'opez, and M.A. Rodr\'\i guez, in preparation.}
that the spectrum of the HS spin chain of $BC_N$ type can be described in terms of a suitable
generalization of Haldane's \emph{motifs}~\cite{HHTBP92} suggests that a similar description
should also exist for the present chain. Note that, for HS chains of $A_{N}$ type, the existence
of such a description is the key ingredient in the proof of the Gaussian character of their level
density~\cite{EFG10} when the number of sites tends to infinity, which is of importance in the
context of quantum chaos and in the study of the thermodynamic properties of these
chains~\cite{EFG12}. In fact, using Eq.~\eqref{ZBN} we have numerically checked that the level
density of the HS chain of $B_N$ type is approximately Gaussian when $N\gtrsim10$, for a wide
range of values of the parameter $\be$ and the spin degrees of freedom $m$ (see, e.g.,
Fig.~\ref{fig.gauss}). This property of HS spin chains of $B_N$ is a further indication of the
existence of a \emph{motif}-based description of their spectrum, which would make possible a
systematic study of the thermodynamics of these chains along the lines of Ref.~\cite{EFG12}.

\section*{Acknowledgments}
This work was supported in part by Spain's MEC, under grant no.~FIS2011-22566.


\begin{thebibliography}{50}
\expandafter\ifx\csname natexlab\endcsname\relax\def\natexlab#1{#1}\fi
\providecommand{\bibinfo}[2]{#2}
\ifx\xfnm\relax \def\xfnm[#1]{\unskip,\space#1}\fi
\bibitem[{Calogero(1971)}]{Ca71}
\bibinfo{author}{F.~Calogero}, \bibinfo{journal}{J. Math. Phys.}
  \bibinfo{volume}{12} (\bibinfo{year}{1971}) \bibinfo{pages}{419--436}.
\bibitem[{Sutherland(1971)}]{Su71}
\bibinfo{author}{B.~Sutherland}, \bibinfo{journal}{Phys. Rev. A}
  \bibinfo{volume}{4} (\bibinfo{year}{1971}) \bibinfo{pages}{2019--2021}.
\bibitem[{Sutherland(1972)}]{Su72}
\bibinfo{author}{B.~Sutherland}, \bibinfo{journal}{Phys. Rev. A}
  \bibinfo{volume}{5} (\bibinfo{year}{1972}) \bibinfo{pages}{1372--1376}.
\bibitem[{Olshanetsky and Perelomov(1983)}]{OP83}
\bibinfo{author}{M.~A. Olshanetsky}, \bibinfo{author}{A.~M. Perelomov},
  \bibinfo{journal}{Phys. Rep.} \bibinfo{volume}{94} (\bibinfo{year}{1983})
  \bibinfo{pages}{313--404}.
\bibitem[{Haldane(1988)}]{Ha88}
\bibinfo{author}{F.~D.~M. Haldane}, \bibinfo{journal}{Phys. Rev. Lett.}
  \bibinfo{volume}{60} (\bibinfo{year}{1988}) \bibinfo{pages}{635--638}.
\bibitem[{Shastry(1988)}]{Sh88}
\bibinfo{author}{B.~S. Shastry}, \bibinfo{journal}{Phys. Rev. Lett.}
  \bibinfo{volume}{60} (\bibinfo{year}{1988}) \bibinfo{pages}{639--642}.
\bibitem[{Ha and Haldane(1992)}]{HH92}
\bibinfo{author}{Z.~N.~C. Ha}, \bibinfo{author}{F.~D.~M. Haldane},
  \bibinfo{journal}{Phys. Rev. B} \bibinfo{volume}{46} (\bibinfo{year}{1992})
  \bibinfo{pages}{9359--9368}.
\bibitem[{Hikami and Wadati(1993)}]{HW93}
\bibinfo{author}{K.~Hikami}, \bibinfo{author}{M.~Wadati}, \bibinfo{journal}{J.
  Phys. Soc. Jpn.} \bibinfo{volume}{62} (\bibinfo{year}{1993})
  \bibinfo{pages}{469--472}.
\bibitem[{Minahan and Polychronakos(1993)}]{MP93}
\bibinfo{author}{J.~A. Minahan}, \bibinfo{author}{A.~P. Polychronakos},
  \bibinfo{journal}{Phys. Lett. B} \bibinfo{volume}{302} (\bibinfo{year}{1993})
  \bibinfo{pages}{265--270}.
\bibitem[{Polychronakos(1993)}]{Po93}
\bibinfo{author}{A.~P. Polychronakos}, \bibinfo{journal}{Phys. Rev. Lett.}
  \bibinfo{volume}{70} (\bibinfo{year}{1993}) \bibinfo{pages}{2329--2331}.
\bibitem[{Sutherland and Shastry(1993)}]{SS93}
\bibinfo{author}{B.~Sutherland}, \bibinfo{author}{B.~S. Shastry},
  \bibinfo{journal}{Phys. Rev. Lett.} \bibinfo{volume}{71}
  (\bibinfo{year}{1993}) \bibinfo{pages}{5--8}.
\bibitem[{Frahm(1993)}]{Fr93}
\bibinfo{author}{H.~Frahm}, \bibinfo{journal}{J. Phys. A: Math. Gen.}
  \bibinfo{volume}{26} (\bibinfo{year}{1993}) \bibinfo{pages}{L473--L479}.
\bibitem[{Polychronakos(1994)}]{Po94}
\bibinfo{author}{A.~P. Polychronakos}, \bibinfo{journal}{Nucl. Phys. B}
  \bibinfo{volume}{419} (\bibinfo{year}{1994}) \bibinfo{pages}{553--566}.
\bibitem[{Finkel and Gonz{\'a}lez-L\'opez(2005)}]{FG05}
\bibinfo{author}{F.~Finkel}, \bibinfo{author}{A.~Gonz{\'a}lez-L\'opez},
  \bibinfo{journal}{Phys. Rev. B} \bibinfo{volume}{72} (\bibinfo{year}{2005})
  \bibinfo{pages}{174411(6)}.
\bibitem[{Ha(1996)}]{Ha96}
\bibinfo{author}{Z.~N.~C. Ha}, \bibinfo{title}{{Q}uantum {M}any-body {S}ystems
  in one {D}imension}, volume~\bibinfo{volume}{12} of
  \textit{\bibinfo{series}{{A}dvances in {S}tatistical {M}echanics}},
  \bibinfo{publisher}{World Scientific}, \bibinfo{address}{Singapore},
  \bibinfo{year}{1996}.
\bibitem[{Murthy and Shankar(1994)}]{MS94}
\bibinfo{author}{M.~V.~N. Murthy}, \bibinfo{author}{R.~Shankar},
  \bibinfo{journal}{Phys. Rev. Lett.} \bibinfo{volume}{73}
  (\bibinfo{year}{1994}) \bibinfo{pages}{3331--3334}.
\bibitem[{Polychronakos(2006)}]{Po06}
\bibinfo{author}{A.~P. Polychronakos}, \bibinfo{journal}{J. Phys. A: Math.
  Gen.} \bibinfo{volume}{39} (\bibinfo{year}{2006})
  \bibinfo{pages}{12793--12845}.
\bibitem[{Azuma and Iso(1994)}]{AI94}
\bibinfo{author}{H.~Azuma}, \bibinfo{author}{S.~Iso}, \bibinfo{journal}{Phys.
  Lett. B} \bibinfo{volume}{331} (\bibinfo{year}{1994})
  \bibinfo{pages}{107--113}.
\bibitem[{Beenakker and Rajaei(1994)}]{BR94}
\bibinfo{author}{C.~W.~J. Beenakker}, \bibinfo{author}{B.~Rajaei},
  \bibinfo{journal}{Phys. Rev. B} \bibinfo{volume}{49} (\bibinfo{year}{1994})
  \bibinfo{pages}{7499--7510}.
\bibitem[{Caselle(1995)}]{Ca95}
\bibinfo{author}{M.~Caselle}, \bibinfo{journal}{Phys. Rev. Lett.}
  \bibinfo{volume}{74} (\bibinfo{year}{1995}) \bibinfo{pages}{2776--2779}.
\bibitem[{Beisert et~al.(2003)Beisert, Kristjansen, and Staudacher}]{BKS03}
\bibinfo{author}{N.~Beisert}, \bibinfo{author}{C.~Kristjansen},
  \bibinfo{author}{M.~Staudacher}, \bibinfo{journal}{Nucl. Phys. B}
  \bibinfo{volume}{664} (\bibinfo{year}{2003}) \bibinfo{pages}{131--184}.
\bibitem[{Beisert(2004)}]{Be04}
\bibinfo{author}{N.~Beisert}, \bibinfo{journal}{Nucl. Phys. B}
  \bibinfo{volume}{682} (\bibinfo{year}{2004}) \bibinfo{pages}{487--520}.
\bibitem[{Bargheer et~al.(2009)Bargheer, Beisert, and Loebbert}]{BBL09}
\bibinfo{author}{T.~Bargheer}, \bibinfo{author}{N.~Beisert},
  \bibinfo{author}{F.~Loebbert}, \bibinfo{journal}{J. Phys. A: Math. Theor.}
  \bibinfo{volume}{42} (\bibinfo{year}{2009}) \bibinfo{pages}{285205(58)}.
\bibitem[{Rej(2012)}]{Re12}
\bibinfo{author}{A.~Rej}, \bibinfo{journal}{Lett. Math. Phys.}
  \bibinfo{volume}{99} (\bibinfo{year}{2012}) \bibinfo{pages}{85--102}.
\bibitem[{Taniguchi et~al.(1995)Taniguchi, Shastry, and Altshuler}]{TSA95}
\bibinfo{author}{N.~Taniguchi}, \bibinfo{author}{B.~S. Shastry},
  \bibinfo{author}{B.~L. Altshuler}, \bibinfo{journal}{Phys. Rev. Lett.}
  \bibinfo{volume}{75} (\bibinfo{year}{1995}) \bibinfo{pages}{3724--3727}.
\bibitem[{Forrester(1994)}]{Fo94}
\bibinfo{author}{P.~J. Forrester}, \bibinfo{journal}{Nucl. Phys. B}
  \bibinfo{volume}{416} (\bibinfo{year}{1994}) \bibinfo{pages}{377--385}.
\bibitem[{van Diejen(1997)}]{Di97}
\bibinfo{author}{J.~F. van Diejen}, \bibinfo{journal}{Commun. Math. Phys.}
  \bibinfo{volume}{188} (\bibinfo{year}{1997}) \bibinfo{pages}{467--497}.
\bibitem[{Dunkl(1998)}]{Du98}
\bibinfo{author}{C.~F. Dunkl}, \bibinfo{journal}{Commun. Math. Phys.}
  \bibinfo{volume}{197} (\bibinfo{year}{1998}) \bibinfo{pages}{451--487}.
\bibitem[{Finkel et~al.(2001)Finkel, G\'omez-Ullate, Gonz{\'a}lez-L\'opez,
  Rodr{{\'\i}}guez, and Zhdanov}]{FGGRZ01}
\bibinfo{author}{F.~Finkel}, \bibinfo{author}{D.~G\'omez-Ullate},
  \bibinfo{author}{A.~Gonz{\'a}lez-L\'opez}, \bibinfo{author}{M.~A.
  Rodr{{\'\i}}guez}, \bibinfo{author}{R.~Zhdanov}, \bibinfo{journal}{Commun.
  Math. Phys.} \bibinfo{volume}{221} (\bibinfo{year}{2001})
  \bibinfo{pages}{477--497}.
\bibitem[{Bernard et~al.(1993)Bernard, Gaudin, Haldane, and Pasquier}]{BGHP93}
\bibinfo{author}{D.~Bernard}, \bibinfo{author}{M.~Gaudin},
  \bibinfo{author}{F.~D.~M. Haldane}, \bibinfo{author}{V.~Pasquier},
  \bibinfo{journal}{J. Phys. A: Math. Gen.} \bibinfo{volume}{26}
  (\bibinfo{year}{1993}) \bibinfo{pages}{5219--5236}.
\bibitem[{Hikami(1995)}]{Hi95npb}
\bibinfo{author}{K.~Hikami}, \bibinfo{journal}{Nucl. Phys. B}
  \bibinfo{volume}{441} (\bibinfo{year}{1995}) \bibinfo{pages}{530--548}.
\bibitem[{Basu-Mallick et~al.(2007)Basu-Mallick, Bondyopadhaya, Hikami, and
  Sen}]{BBHS07}
\bibinfo{author}{B.~Basu-Mallick}, \bibinfo{author}{N.~Bondyopadhaya},
  \bibinfo{author}{K.~Hikami}, \bibinfo{author}{D.~Sen},
  \bibinfo{journal}{Nucl. Phys. B} \bibinfo{volume}{782} (\bibinfo{year}{2007})
  \bibinfo{pages}{276--295}.
\bibitem[{Beisert and Erkal(2008)}]{BE08}
\bibinfo{author}{N.~Beisert}, \bibinfo{author}{D.~Erkal}, \bibinfo{journal}{J.
  Stat. Mech.} \bibinfo{volume}{0803} (\bibinfo{year}{2008})
  \bibinfo{pages}{P03001}.
\bibitem[{Bernard et~al.(1995)Bernard, Pasquier, and Serban}]{BPS95}
\bibinfo{author}{D.~Bernard}, \bibinfo{author}{V.~Pasquier},
  \bibinfo{author}{D.~Serban}, \bibinfo{journal}{Europhys. Lett.}
  \bibinfo{volume}{30} (\bibinfo{year}{1995}) \bibinfo{pages}{301--306}.
\bibitem[{Yamamoto(1995)}]{Ya95}
\bibinfo{author}{T.~Yamamoto}, \bibinfo{journal}{Phys. Lett. A}
  \bibinfo{volume}{208} (\bibinfo{year}{1995}) \bibinfo{pages}{293--302}.
\bibitem[{Yamamoto and Tsuchiya(1996)}]{YT96}
\bibinfo{author}{T.~Yamamoto}, \bibinfo{author}{O.~Tsuchiya},
  \bibinfo{journal}{J. Phys. A: Math. Gen.} \bibinfo{volume}{29}
  (\bibinfo{year}{1996}) \bibinfo{pages}{3977--3984}.
\bibitem[{Corrigan and Sasaki(2002)}]{CS02}
\bibinfo{author}{E.~Corrigan}, \bibinfo{author}{R.~Sasaki},
  \bibinfo{journal}{J. Phys. A: Math. Gen.} \bibinfo{volume}{35}
  (\bibinfo{year}{2002}) \bibinfo{pages}{7017--7061}.
\bibitem[{Finkel et~al.(2003)Finkel, G\'omez-Ullate, Gonz{\'a}lez-L\'opez,
  Rodr{{\'\i}}guez, and Zhdanov}]{FGGRZ03}
\bibinfo{author}{F.~Finkel}, \bibinfo{author}{D.~G\'omez-Ullate},
  \bibinfo{author}{A.~Gonz{\'a}lez-L\'opez}, \bibinfo{author}{M.~A.
  Rodr{{\'\i}}guez}, \bibinfo{author}{R.~Zhdanov}, \bibinfo{journal}{Commun.
  Math. Phys.} \bibinfo{volume}{233} (\bibinfo{year}{2003})
  \bibinfo{pages}{191--209}.
\bibitem[{Enciso et~al.(2005)Enciso, Finkel, Gonz{\'a}lez-L\'opez, and
  Rodr{{\'\i}}guez}]{EFGR05}
\bibinfo{author}{A.~Enciso}, \bibinfo{author}{F.~Finkel},
  \bibinfo{author}{A.~Gonz{\'a}lez-L\'opez}, \bibinfo{author}{M.~A.
  Rodr{{\'\i}}guez}, \bibinfo{journal}{Nucl. Phys. B} \bibinfo{volume}{707}
  (\bibinfo{year}{2005}) \bibinfo{pages}{553--576}.
\bibitem[{Barba et~al.(2008)Barba, Finkel, Gonz\'alez-L\'opez, and
  Rodr{\'\i}guez}]{BFGR08}
\bibinfo{author}{J.~C. Barba}, \bibinfo{author}{F.~Finkel},
  \bibinfo{author}{A.~Gonz\'alez-L\'opez}, \bibinfo{author}{M.~A.
  Rodr{\'\i}guez}, \bibinfo{journal}{Phys. Rev. B} \bibinfo{volume}{77}
  (\bibinfo{year}{2008}) \bibinfo{pages}{214422(10)}.
\bibitem[{Barba et~al.(2009)Barba, Finkel, Gonz\'alez-L\'opez, and
  Rodr{\'\i}guez}]{BFGR09}
\bibinfo{author}{J.~C. Barba}, \bibinfo{author}{F.~Finkel},
  \bibinfo{author}{A.~Gonz\'alez-L\'opez}, \bibinfo{author}{M.~A.
  Rodr{\'\i}guez}, \bibinfo{journal}{Nucl. Phys. B} \bibinfo{volume}{806}
  (\bibinfo{year}{2009}) \bibinfo{pages}{684--714}.
\bibitem[{Khastgir et~al.(2000)Khastgir, Pocklington, and Sasaki}]{KPS00}
\bibinfo{author}{S.~P. Khastgir}, \bibinfo{author}{A.~J. Pocklington},
  \bibinfo{author}{R.~Sasaki}, \bibinfo{journal}{J. Phys. A: Math. Gen.}
  \bibinfo{volume}{33} (\bibinfo{year}{2000}) \bibinfo{pages}{9033--9064}.
\bibitem[{Loris and Sasaki(2004)}]{LS04}
\bibinfo{author}{I.~Loris}, \bibinfo{author}{R.~Sasaki}, \bibinfo{journal}{J.
  Phys. A: Math. Gen.} \bibinfo{volume}{37} (\bibinfo{year}{2004})
  \bibinfo{pages}{211--237}.
\bibitem[{Basu-Mallick et~al.(2009)Basu-Mallick, Finkel, and
  Gonz{\'a}lez-L{\'o}pez}]{BFG09}
\bibinfo{author}{B.~Basu-Mallick}, \bibinfo{author}{F.~Finkel},
  \bibinfo{author}{A.~Gonz{\'a}lez-L{\'o}pez}, \bibinfo{journal}{Nucl. Phys. B}
  \bibinfo{volume}{812} (\bibinfo{year}{2009}) \bibinfo{pages}{402--423}.
\bibitem[{Basu-Mallick et~al.(2011)Basu-Mallick, Finkel, and
  Gonz{\'a}lez-L{\'o}pez}]{BFG11}
\bibinfo{author}{B.~Basu-Mallick}, \bibinfo{author}{F.~Finkel},
  \bibinfo{author}{A.~Gonz{\'a}lez-L{\'o}pez}, \bibinfo{journal}{Nucl. Phys. B}
  \bibinfo{volume}{843} (\bibinfo{year}{2011}) \bibinfo{pages}{505--553}.
\bibitem[{Simon(1983)}]{Si83}
\bibinfo{author}{B.~Simon}, \bibinfo{journal}{Ann. Inst. H. Poincar{\'e} Sect.
  A (N. S.)} \bibinfo{volume}{38} (\bibinfo{year}{1983})
  \bibinfo{pages}{295--308}.
\bibitem[{Ahmed et~al.(1979)Ahmed, Bruschi, Calogero, Olshanetsky, and
  Perelomov}]{ABCOP79}
\bibinfo{author}{S.~Ahmed}, \bibinfo{author}{M.~Bruschi},
  \bibinfo{author}{F.~Calogero}, \bibinfo{author}{M.~A. Olshanetsky},
  \bibinfo{author}{A.~M. Perelomov}, \bibinfo{journal}{Nuovo Cimento B}
  \bibinfo{volume}{49} (\bibinfo{year}{1979}) \bibinfo{pages}{173--199}.
\bibitem[{Haldane et~al.(1992)Haldane, Ha, Talstra, Bernard, and
  Pasquier}]{HHTBP92}
\bibinfo{author}{F.~D.~M. Haldane}, \bibinfo{author}{Z.~N.~C. Ha},
  \bibinfo{author}{J.~C. Talstra}, \bibinfo{author}{D.~Bernard},
  \bibinfo{author}{V.~Pasquier}, \bibinfo{journal}{Phys. Rev. Lett.}
  \bibinfo{volume}{69} (\bibinfo{year}{1992}) \bibinfo{pages}{2021--2025}.
\bibitem[{Enciso et~al.(2010)Enciso, Finkel, and Gonz{\'a}lez-L\'opez}]{EFG10}
\bibinfo{author}{A.~Enciso}, \bibinfo{author}{F.~Finkel},
  \bibinfo{author}{A.~Gonz{\'a}lez-L\'opez}, \bibinfo{journal}{Phys. Rev. E}
  \bibinfo{volume}{82} (\bibinfo{year}{2010}) \bibinfo{pages}{051117(6)}.
\bibitem[{Enciso et~al.(2012)Enciso, Finkel, and Gonz{\'a}lez-L\'opez}]{EFG12}
\bibinfo{author}{A.~Enciso}, \bibinfo{author}{F.~Finkel},
  \bibinfo{author}{A.~Gonz{\'a}lez-L\'opez}, \bibinfo{title}{Thermodynamics of
    spin chains of {H}aldane--{S}hastry type and one-dimensional vertex models},
  \bibinfo{journal}{Ann. Phys.-New York,}
  \bibinfo{year}{in press}. \bibinfo{note}{ArXiv:1204.3805v1 [cond-mat.stat-mech]}.

\end{thebibliography}

\end{document}